\documentclass[11pt]{article}
\usepackage{url}

\usepackage{hyperref}

\usepackage{amssymb,amscd,amstext,amsmath, amsthm, graphicx}
\usepackage{latexsym}

\usepackage{floatflt}
\usepackage[mathscr]{eucal}
\usepackage{tikz}
\usetikzlibrary{arrows}
\usepackage{verbatim}
\usepackage{soul}
\usepackage{wrapfig}
\usepackage{fancyhdr}
\usepackage{hyperref}
\usepackage{setspace}
\usepackage{mathdots}
\usepackage{blindtext}
\usepackage{enumitem}
 \usepackage{placeins}
\usepackage[top=1in, bottom=1in, left=1in, right=1in]{geometry}

\usepackage{setspace}

\definecolor{linkcolor}{HTML}{000000}
\definecolor{urlcolor}{HTML}{000000}
\hypersetup{pdfstartview=Fit, linkcolor=linkcolor,urlcolor=urlcolor,
colorlinks=true, citecolor=linkcolor}
\renewenvironment{abstract}{%
\begin{center}\begin{minipage}{0.9\textwidth}\begin{small}
\textbf{Abstract.}}
{\end{small}\par\noindent\end{minipage}\end{center}}

\title{\bf Problems and solutions of the Fourth International Students' Olympiad in Cryptography NSUCRYPTO\footnote{The paper was supported by the Russian Ministry of Science and Education (the 5-100 Excellence Programme and the Project no. 1.12875.2018/12.1), by the Russian Foundation for Basic Research (projects no. 17-41-543364, 18-07-01394, 18-31-00479, 18-31-00374), by the program of fundamental scientific researches of the SB RAS no.I.5.1. (project no. 0314-2016-0017).}
}
\author{A.~Gorodilova$^{1,2}$,
    S.~Agievich$^{3}$,
    C.~Carlet$^{4}$,
    E.~Gorkunov$^{1,2}$,
    V.~Idrisova$^{1,2}$,\\
    N.~Kolomeec$^{1,2}$,
    A.~Kutsenko$^{1}$,
    S.~Nikova$^{5}$,
    A.~Oblaukhov$^{1}$,
    S.~Picek$^{6}$,\\
    B.~Preneel$^{5}$,
    V.~Rijmen$^{5}$,
    N.~Tokareva$^{1,2}$
    \\
  \\
    {\small $^{1}$Novosibirsk State University, Novosibirsk, Russia}\\
    {\small $^{2}$Sobolev Institute of Mathematics, Novosibirsk, Russia} \\
  {\small$^{3}$Belarusian State University, Minsk, Belarus} \\
  {\small$^{4}$University of Paris 8, Paris, France} \\
  {\small$^{5}$ESAT-COSIC, KU Leuven, Leuven, Belgium} \\
  {\small$^{6}$Cyber Security Research Group, Delft University of Technology, Mekelweg 2, Delft, The Netherlands} \\  
    \\
    {\small E-mail: {\tt nsucrypto@nsu.ru}}
    }
\date{}

\begin{document}

\pagenumbering{arabic}

\maketitle

\begin{abstract}
Mathematical problems and their solutions of the Fourth International Students' Olympiad in cryptography NSUCRYPTO'2017 are presented. We consider problems related to attacks on ciphers and hash functions, cryptographic Boolean functions, the linear branch number, addition chains, error correction codes, etc. We discuss several open problems on algebraic structure of cryptographic functions, useful proof-of-work algorithms, the Boolean hidden shift problem and quantum computings.

\vspace{0.2cm}

\end{abstract}

\section*{Introduction}

The International Students' Olympiad in cryptography NSUCRYPTO was held for the fourth time in 2017. The idea of the Olympiad was born in Novosibirsk State University that is located in the world-famous scientific center in the heart of Siberia --- Akademgorodok. Now, the Olympiad program committee includes specialists from Belgium, France, The Netherlands, USA, Norway, India, Belarus', Russia. At the same time the geography of participants is expanding year by year: there were more than 1300 participants from 38 countries in 2014--2017.

Let us shortly formulate the format of the Olympiad (all information
can be found on the official website
\href{https://nsucrypto.nsu.ru/}{\textcolor{blue}{\tt
nsucrypto.nsu.ru}}). One of the Olympiad ideas is that everyone can
participate: school students, university students and professionals!
Each participant chooses his/her category when registering on the
Olympiad website. The Olympiad consists of two independent Internet
rounds: the first one is individual while the second round is team.
The first round is divided into two sections: A --- for school
students, B --- for university students and professionals.
Participants read the Olympiad problems and send their solutions
using the Olympiad website. The language of the Olympiad is English.

Every year, participants are offered to solve several problems of
different complexity at the intersection of mathematics and
cryptography. Another feature of the Olympiad is that it not only
includes interesting tasks with known solutions but also offers
unsolved problems in this area. 
All the open problems stated during
the Olympiad history can be found at\\
\href{https://nsucrypto.nsu.ru/unsolved-problems}{\textcolor{blue}{\tt
nsucrypto.nsu.ru/unsolved-problems}}. On the website we also mark
the current status of each problem. For example, the problem
``Algebraic immunity'' (2016) was completely solved during the
Olympiad, a partial solution for the problem ``A secret sharing''
(2014) was proposed in \cite{sharing}. We invite everybody who has
ideas how to solve the problems to send your solutions to~us!

What was surprising and very pleasant for us this year is that the NSUCRYPTO Olympiad can even change the professional life of people! Following are some examples of the feedback we have received over the years:

``..{\sl Without having a 3rd ranking to go for PhD, I would give up in cryptography due to the complex mathematic involve}..'' --- Duc Tri Nguyen (Vietnam, 3rd places in 2016 and 2017);
``..{\sl When we join the competition, we just want to know where we are with our knowledge, want to test ourself with our self-study Crypto, and want to improve our math in Cryptography}..'' --- Quan Doan (Vietnam, 3rd places in 2016 and 2017);
``..{\sl The problems contain much knowledge not only from mathematics and cryptography, but also many other fields such as art..}'' --- Renzhang Liu (China, 1st place in 2015);
``{\sl ..I sometimes find myself reading too much and this competition is a great way of putting knowledge into practice by solving fun tasks}..'' --- Dragos Alin Rotaru (United Kingdom, 3d place in 2016);
``{\sl ..Its a contest that leaves you wanting to spend more time on it after the deadline, just to work out the questions you didn't get..}'' --- Robert Spencer (United Kingdom, 1st place in 2016, 2d place in 2017);
``{\sl ..Participation in the Olympiad offers you an excellent opportunity to try yourself as a codebreaker and a cryptographer..}'' --- Anna Taranenko (Russia, 1st place in 2014, 2d places in 2015 and 2016, honorable diploma in 2017);
``{\sl ..Sometimes you spend hours trying to solve a problem, sometimes it seems that it's impossible to solve it. But when you find a solution, usually so obvious, you experience an incomparable sense of delight. My hat's off to the person who think up such interesting tasks!..}'' --- Evgeniya Ishchukova (Russia, 3d place in 2016, honorable diplomas in 2015 and 2017). A complete list of comments can be found at \href{https://nsucrypto.nsu.ru/feedbacks}{\textcolor{blue}{\tt nsucrypto.nsu.ru/feedbacks}}.

The paper is organized as follows. We start with problem structure of the Olympiad in section~\ref{problem-structure}. Then we present formulations of all the problems stated during the Olympiad and give their detailed solutions (section~\ref{problems}). Finally, we publish the lists of NSUCRYPTO'2017 winners in section \ref{winners}.

Mathematical problems of the previous International Olympiads
NSUCRYPTO'2014, NSUCRYPTO'2015, and NSUCRYPTO'2016 can be found in \cite{paper-pdm}, \cite{paper-cryptologia}, and \cite{paper-PDM-18} respectively.

\section{Problem structure of the Olympiad}
\label{problem-structure}

There were 16 problems stated during the Olympiad, some of them were included
in both rounds (Tables\;\ref{Probl-First},\,\ref{Probl-Second}).
The school section of the first round consisted of six problems, whereas the
student section contained seven problems. Two problems were common for both sections.
The second round was composed of eleven problems; they were common for all the
participants. Three problems of the second round were marked as
unsolved (awarded special prizes from the Program Committee).

\begin{table}[ht]
\centering\footnotesize
\caption{Problems of the first round}
\medskip
\label{Probl-First}
\begin{tabular}{cc}
\begin{tabular}{|c|l|c|}
  \hline
  N & Problem title & Maximum scores \\
  \hline
  \hline
  1 & PIN code & 4 \\
    \hline
  2 & Chests with treasure & 4 \\
    \hline
  3 & A numerical rebus & 4 \\
    \hline
  4 & Timing attack & 4 \\
    \hline
  5 & The shortest addition chain & 4 \\
    \hline
  6 & A music lover & 4 \\
  \hline
\end{tabular}
&
\begin{tabular}{|c|l|c|}
  \hline
  N & Problem title & Maximum scores \\
  \hline
    \hline
  1 & Timing attack & 4 \\
    \hline
  2 & Treasure chests & 4 \\
    \hline
  3 & A music lover & 4 \\
    \hline
  4 & An infinite set of collisions & 4 \\
    \hline
  5 & One more parameter & 10 \\
    \hline
  6 & Scientists & 8 \\
  \hline
  7 & Masking & 10 \\
  \hline
\end{tabular}
\\
\noalign{\smallskip}
A --- school section
&
B --- student section
\\
\end{tabular}
\end{table}

\begin{table}[ht]
\centering\footnotesize
\caption{Problems of the second round}
\smallskip
\label{Probl-Second}
\begin{tabular}{|c|l|c|}
  \hline
  N & Problem title & Maximum scores \\
  \hline
    \hline
  1 & The image set & Unsolved \\
    \hline
  2 & {\tt TwinPeaks} & 8 \\
    \hline
  3 & An addition chain & 8 \\
    \hline
  4 & Hash function {\tt FNV2} & 8 \\
    \hline
  5 & A music lover & 4 \\
    \hline
  6 & Boolean hidden shift and quantum computings & Unsolved  \\
    \hline
  7 & One more parameter & 10 \\
    \hline
  8 & Scientists & 8 \\
    \hline
  9 & Masking & 10 \\
    \hline
  10 & PIN code & 4 \\
    \hline
  11 & Useful proof-of-work for blockchains & Unsolved \\
  \hline
\end{tabular}
\end{table}

\section{Problems and their solutions}
\label{problems}
In this section we formulate all the problems of NSUCRYPTO'2017 and present their detailed solutions with paying attention to solutions proposed by the participants.

\subsection{Problem ``PIN code''}

\subsubsection{Formulation}\label{pincode}

A PIN code $P=\overline{p_1 p_2 \ldots p_n}$\ \  is an arbitrary number consisting of a finite number of pairwise different decimal digits in ascending order ($p_1 < p_2 < \ldots < p_n$). Bob got his personal PIN code in the bank, but he decided that the code is not secure enough and changed it in the following~way:
\begin{itemize}[noitemsep]
 \item[{\bf 1}.] Bob multiplied his PIN code $P$ by 999 and obtained the number $A=\overline{a_1 a_2 \ldots a_m}$;
 \item[{\bf 2}.]  Then he found the sum of all digits of $A$: $a_1+a_2+\ldots + a_m=S=\overline{s_1 s_2 \ldots s_k}$;

 \item[{\bf 3}.] Finally, he took all digits (starting from 0) that are smaller than $s_1$, sorted them in ascending order and inserted between digits $s_1$ and $s_2$ in the number $S$. The resulting number $P'$ is Bob's new PIN code.
For example, if $S$ was $345$, then, after such insertion we obtain $P' = 301245$.
\end{itemize}

Find the new code $P'$!

\textbf{Remarks}. By $\overline{p_1p_2\ldots p_n}$ we mean that $p_1, p_2, \ldots,p_n$ are decimal digits and all digits over the bar form a decimal number.

\subsubsection{Solution}

Let $P=\overline{p_1 p_2 \ldots p_n}$ for some positive integer $n$.
Let us note that $P$ multiplied by $999$ is the same thing as $P$ multiplied by 1000 minus $P $, that is shifting the number $P$ on three positions to the left minus itself. Let us consider this subtraction:
 \begin{center}
$ \begin{matrix}
& p_1 & p_2 & p_3 & p_4 & \cdots & p_n  & 0 &0&0\\
   -& &  & & p_1 & \cdots & p_{n-3} & p_{n-2} & p_{n-1}  &p_n
 \end{matrix}
 $
 \end{center}
Since $p_1 < p_2 < \ldots < p_n$ by definition, we have that $p_{n-3} < p_n$. Therefore, we borrow a unit only from $p_n$ among $p_1,\ldots,p_n$. Thus, the sum of digits of this difference is equal to $$(10-p_n)+(9-p_{n-1})+(9-p_{n-2})+(p_{n}-1-p_{n-3})+(p_{n-1}-p_{n-4})\ldots + (p_4-p_1)+p_3+p_2+p_1=27.$$

So, Bob's new PIN code is $2017$.

 A lot of great and compact solutions were sent by the participants. The best solution in the school section was by  Lenart Bucar (Gimnazija Be\v{z}igrad, Grosuplje, Slovenia). Also we want to note a detailed solution sent by Ivan Baksheev (Gymnasium 6, Novosibirsk, Russia).

\subsection{Problem ``Chests with treasure''}

\subsubsection{Formulation}

We have three closed chests. Some of them contain treasures
(diamonds, gold coins, bitcoins as well) but we do not know which
ones. A parrot knows which chests contain treasures and which do
not; he agrees to answer to questions with ``yes'' or ``no''. He may
possibly lie in his answers but not more than once. List six
questions such that it is possible to deduce from the answers of the
parrot which chests contain treasures and which do not.

\subsubsection{Solution}\label{chests}
Let us mark the chest with 0 if it does not contain the treasure, and with 1 if it does. Similarly, let us write down parrot's answers to our questions as 1's (for ``yes'') and 0's (for ``no''). Now the state of chests is encoded with three bits of information and parrot's answers give us six bits of information. Taking into account that the parrot may or may not lie in one of his answers, one or none of the bits among these six can be faulty.

So, our goal is to devise six questions in such a way, that six bits of obtained information can be uniquely decoded to exactly one 3-bit state of chests, even if one bit of answers was not right. In terms of coding theory, we need to construct an error-correcting code, which corrects one error. In simple terms, we need to map every one of $2^3=8$ chest states to a Boolean vector of length six in such a way, that even if one bit in any of these image vectors gets flipped, we can uniquely determine which vector it was before the error occurred.

To find such eight vectors it is sufficient to find eight vectors of length six, such that every two of them are different in at least three bits. Even if we receive one of these vectors with a flipped bit, it will be possible to determine which of these eight vectors it was before the error. Such eight vectors can be found manually, one example is shown below:

\begin{center}
\begin{tabular}{llllllll}

$(000)\rightarrow (000000)$
& ~ &
$(010)\rightarrow (110011)$
& ~ &
$(011)\rightarrow (100101)$
& ~ &
$(110)\rightarrow (010110)$
\\
$(001)\rightarrow (001111)$
& ~ &
$(100)\rightarrow (111100)$
& ~ &
$(101)\rightarrow (011001)$
& ~ &
$(111)\rightarrow (101010)$\\
\end{tabular}
\end{center}

So, if we are able to make a list of six questions, which for every 3-bit state of the chests gives us corresponding 6-bit answer vector, we will be able to reconstruct the correct answer vector even if the parrot lies when answering one of the questions.

The questions can easily be constructed by using logical operations. Let us take a look at the first bit of all answer vectors. It is equal to 1 only in 3rd, 4th, 5th and 8th vectors, corresponding to chest states $(010),(100),(011),(111)$. So we can ask the first question like this:

\begin{center}
``Is this true that there are treasures {\bf only in the {\bf second} chest} or {\bf only in the {\bf first} chest} or\\ {\bf only in {\bf second and third} chests} or {\bf in {\bf all} chests}?''
\end{center}

The truthful answer to this question will give us the correct first bit of an answer vector. By constructing other five questions similarly, we can create a mapping described above, and can be able to decode six answers into 3-bit chest state even if one answer was not honest.

Originally, it was intended that the list of six questions should be presented before asking any of them. So, all questions are predetermined and do not depend on answers for other questions. But since this was not explicitly stated in the description of the problem, this condition was not required.
Correct and full solutions to the problem were given by seven participants, most of them do not use such a ``coding theory'' approach and present interesting questioning strategies for the problem. The best solutions were presented by Alexander Grebennikov (Presidential PML 239, St. Petersburg, Russia) and by Ivan Baksheev (Gymnasium 6, Novosibirsk, Russia).

\subsection{Problem ``Treasure chests''}

\subsubsection{Formulation}

We have seven closed chests. Some of them contain treasures
(diamonds, gold coins, bitcoins as well) but we do not know which
ones. A parrot knows which chests contain treasures and which do
not; he agrees to answer to questions with ``yes'' or ``no''. But he
may possibly lie in his answers but not more than twice. List
fifteen questions such that it is possible to deduce from the
answers of the parrot which chests contain treasures and which
do not.

\subsubsection{Solution}  See solution to the problem ``Chests with treasure'' (section \ref{chests}) for the idea of the solution.
In this problem we need to find a code, which maps all binary vectors of length seven (chest states) to binary vectors of length fifteen and can correct up to two errors. If such a code is found, then we can easily construct questions using technique described in the solution for the ``Chests with treasure'' problem.

It is sufficient to find a $[15,7,5]$-code, where ``5'' is the minimal distance of the code, since a code with the minimal distance five or greater can correct two errors.
It is known that there exist BCH (Bose~---~Chaudhuri~---~Hocquenghem) codes with these exact parameters (for example, see \cite{MacW-Sl}). Using one of those, we can construct fifteen questions for the parrot which will allow us to decode chest states from answers.

Full and correct solutions for the problem were proposed by twelve university students and professionals. All of them used error correction codes.

\subsection{Problem ``A numerical rebus''}

\subsubsection{Formulation}

Buratino keeps his Golden Key in a safe that is locked with a numerical password.
For secure storage of the password he replaced some digits in the password by letters (in a way that different letters substitute different digits). After replacement Buratino got the password
{\bf NSUCRYPTO17}.
Alice the Fox found out that:
\begin{itemize}[noitemsep]
\item the number \textbf{NSUCRYPTO} is divisible by all integers $n$, where $n<17$, and
\item the difference $\textbf{NSU $-$ CRY}$ is divisible by 7.
\end{itemize}

Can she find the password?

\medskip

\textbf{Remarks}. Here we denote $\overline{ABC\ldots}$ (see section \ref{pincode}) by \textbf{ABC}$\ldots$.

\subsubsection{Solution}  The main idea of the solution is to apply necessary divisibility rules that would allow us to reduce the exhaustive search for the original password. Let us describe the main steps of the solution.
\begin{itemize}[noitemsep]
 \item[{\bf 1}.] Since the number {\bf NSUCRYPTO} is divisible by all integers $1,2,\ldots,16$, it is divisible by 720720. Thus, {\bf O} is equal to 0.

 \item[{\bf 2}.] Since {\bf NSU} $-$ {\bf CRY} is divisible by 7, {\bf PTO} is also divisible by 7.

 \item[{\bf 3}.] Since {\bf NSUCRYPTO} is divisible by 8, {\bf PTO} is divisible by 8. Thus, we have that {\bf PTO} is equal to 280, 560, or 840.

 \item[{\bf 4}.] Since {\bf O} is equal to 0 and {\bf NSUCRYPTO} is divisible by 9, there is no digit 9 in this number.

 \item[{\bf 5}.] Then we check all the possible variants, regarding divisibility rules for 11 and 13. Finally, we find the unique number that is 376215840. So, the password is 37621584017.
\end{itemize}
The best solution was proposed by Andrei Razvan (``Fratii Buzesti'' National College, Craiova, Romania). He implemented steps 1--3 and searched through the rest of possible numbers using a computer. While many proposed solutions used exhaustive searches, many students tried to find this number theoretically and got partial results.

\subsection{Problem ``Timing attack''}

\subsubsection{Formulation}
 Anton invented a ciphermachine that can automatically encrypt messages consisting of English letters. Each letter corresponds to the number from 1 to 26 by alphabetical order (1 is for {\tt A}, 2 is for {\tt B}, ..., 26 is for {\tt Z}). The machine encrypts messages letter by letter. It encrypts one letter as follows.
\begin{itemize}[noitemsep]
\item[{\bf1.}] If the letter belongs to the special secret set of letters, the machine does not encrypt it, adds the original letter to the ciphertext, and does not go to step~2; otherwise it goes to step~2.
\item[{\bf2.}] According to the secret rule, it replaces the current letter with number $k$ by a letter with number $\ell$, where $\ell$ has the same remainder of division by 7, and adds this new letter to the ciphertext.
\end{itemize}

 Anton's classmate Evgeny is interested in different kinds of cryptanalysis that use some physical information about the encryption process. He measured the amount of time that is required for each letter encryption by Anton's ciphermachine and found out that a timing attack can be applied to it!

  He captured the ciphertext that Anton sent to his friend and was able to read the message using the information of his measurements!

Could you also decrypt the ciphertext

\medskip

{\tt Tois kevy is fhye tvvu xust hgvtoed iyife ngfbey! }

{\tt Wvat ka rvn knvw owvnt it?}

\begin{figure}[h!]
\centering
\includegraphics[width=0.7\textwidth]{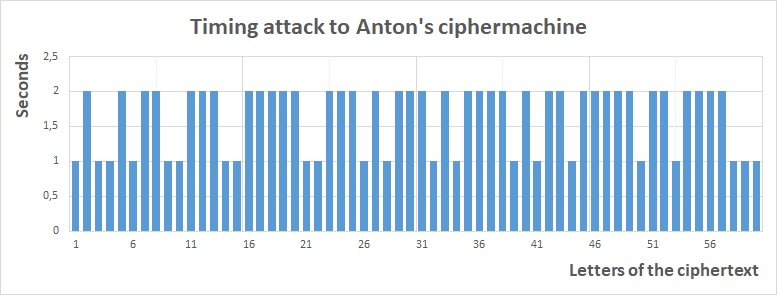}
\vspace{-2mm}
\caption{{\small Time consumption of the message encryption}}
\label{tim-at}
\end{figure}

\medskip

\noindent if you know how much time encryption of each message letter took (Figure\,\ref{tim-at})?

\FloatBarrier

\subsubsection{Solution}  First of all one can notice that according to the diagram (Figure\,\ref{tim-at}) an encryption of each letter took one or two seconds. It leads to an idea that the encryption process takes one second for letters from secret group (only step 1 is needed) and takes two seconds otherwise (2 steps of encryption). Thus, one can easily find letters that belong the secret group and partially read the message:

\medskip

{\tt T\_is \_e\_\_ is \_\_\_e t\_\_\_ \_\_st \_\_\_t\_e\_ \_\_i\_e \_\_\_\_e\_!}

{\tt W\_\_t \_\_ \_\_\_ k\_\_w \_\_\_\_t it?}

\medskip

To read the whole message we just need to write all possible replacements for each of empty positions according to the step 2 rule and to find the appropriate English words. Note that the second sentence ``{\tt W\b{h}\b{a}t \b{d}\b{o} \b{y}\b{o}\b{u} k\b{n}\b{o}w \b{a}\b{b}\b{o}\b{u}t it?}'' can be correctly guessed at the first glance! And many participants mentioned this observation. Let us read the first sentence:

\renewcommand{\arraystretch}{0.7}
\renewcommand{\tabcolsep}{0.4mm}

\begin{center}
{\tt
\begin{tabular}{*{49}{c}}
T &
\begin{tabular}[c]{l}
a\\
\hline
\multicolumn{1}{|c|}{h}\\
\hline
o\\
v\\
\end{tabular} &
i &
s &
~  &
\begin{tabular}[c]{l}
d\\
k\\
r\\
\hline
\multicolumn{1}{|c|}{y}\\
\hline
\end{tabular} &
e &
\begin{tabular}[c]{l}
\hline
\multicolumn{1}{|c|}{a}\\
\hline
h\\
o\\
v\\
\end{tabular} &
\begin{tabular}[c]{l}
d\\
k\\
\hline
\multicolumn{1}{|c|}{r}\\
\hline
y\\
\end{tabular} &
~ &
i &
s &
~ &
\begin{tabular}[c]{l}
f\\
\hline
\multicolumn{1}{|c|}{m}\\
\hline
t\\
\end{tabular} &
\begin{tabular}[c]{l}
a\\
h\\
\hline
\multicolumn{1}{|c|}{o}\\
\hline
v\\
\end{tabular} &
\begin{tabular}[c]{l}
d\\
k\\
\hline
\multicolumn{1}{|c|}{r}\\
\hline
y\\
\end{tabular} &
e &
~ &
t &
\begin{tabular}[c]{l}
a\\
\hline
\multicolumn{1}{|c|}{h}\\
\hline
o\\
v\\
\end{tabular} &
\begin{tabular}[c]{l}
\hline
\multicolumn{1}{|c|}{a}\\
\hline
h\\
o\\
v\\
\end{tabular} &
\begin{tabular}[c]{l}
g\\
\hline
\multicolumn{1}{|c|}{n}\\
\hline
u\\
\end{tabular} &
~  &
\begin{tabular}[c]{l}
c\\
\hline
\multicolumn{1}{|c|}{j}\\
\hline
q\\
x\\
\end{tabular} &
\begin{tabular}[c]{l}
g\\
n\\
\hline
\multicolumn{1}{|c|}{u}\\
\hline
\end{tabular} &
s &
t &
~ &
\begin{tabular}[c]{l}
\hline
\multicolumn{1}{|c|}{a}\\
\hline
h\\
o\\
v\\
\end{tabular} &
\begin{tabular}[c]{l}
g\\
\hline
\multicolumn{1}{|c|}{n}\\
\hline
u\\
\end{tabular} &
\begin{tabular}[c]{l}
a\\
h\\
\hline
\multicolumn{1}{|c|}{o}\\
\hline
v\\
\end{tabular} &
t &
\begin{tabular}[c]{l}
a\\
\hline
\multicolumn{1}{|c|}{h}\\
\hline
o\\
v\\
\end{tabular} &
e &
\begin{tabular}[c]{l}
d\\
k\\
\hline
\multicolumn{1}{|c|}{r}\\
\hline
y\\
\end{tabular} &
~ &
\begin{tabular}[c]{l}
b\\
i\\
\hline
\multicolumn{1}{|c|}{p}\\
\hline
w\\
\end{tabular} &
\begin{tabular}[c]{l}
d\\
k\\
\hline
\multicolumn{1}{|c|}{r}\\
\hline
y\\
\end{tabular} &
i &
\begin{tabular}[c]{l}
f\\
\hline
\multicolumn{1}{|c|}{m}\\
\hline
t\\
\end{tabular} &
e &
~ &
\begin{tabular}[c]{l}
g\\
\hline
\multicolumn{1}{|c|}{n}\\
\hline
u\\
\end{tabular} &
\begin{tabular}[c]{l}
g\\
n\\
\hline
\multicolumn{1}{|c|}{u}\\
\hline
\end{tabular} &
\begin{tabular}[c]{l}
f\\
\hline
\multicolumn{1}{|c|}{m}\\
\hline
t\\
\end{tabular} &
\begin{tabular}[c]{l}
\hline
\multicolumn{1}{|c|}{b}\\
\hline
i\\
p\\
w\\
\end{tabular} &
e &
\begin{tabular}[c]{l}
d\\
k\\
\hline
\multicolumn{1}{|c|}{r}\\
\hline
y\\
\end{tabular} &
!
\\
\end{tabular}
}
\end{center}
Thus, the message is ``{\tt This year is more than just another prime number!} {\tt What do you know about it?}''. Really, the number 2017 has magnificent properties (see \cite{2017-prime})!
Almost all participants solved the problem and found the correct answer. The best ones were given by school students Borislav Kirilov (FPMG, Sofia, Bulgaria) and Vladimir Schavelev (SESC NSU,   Novosibirsk, Russia), and  by university students Igor Antonov (Ural State University of Railway Transport) and Kristina   Zhuchenko (Demidov Yaroslavl State University).

\subsection{Problem ``A music lover''}

\subsubsection{Formulation}
\begin{figure}[h!]
\centering
\includegraphics[width=\textwidth]{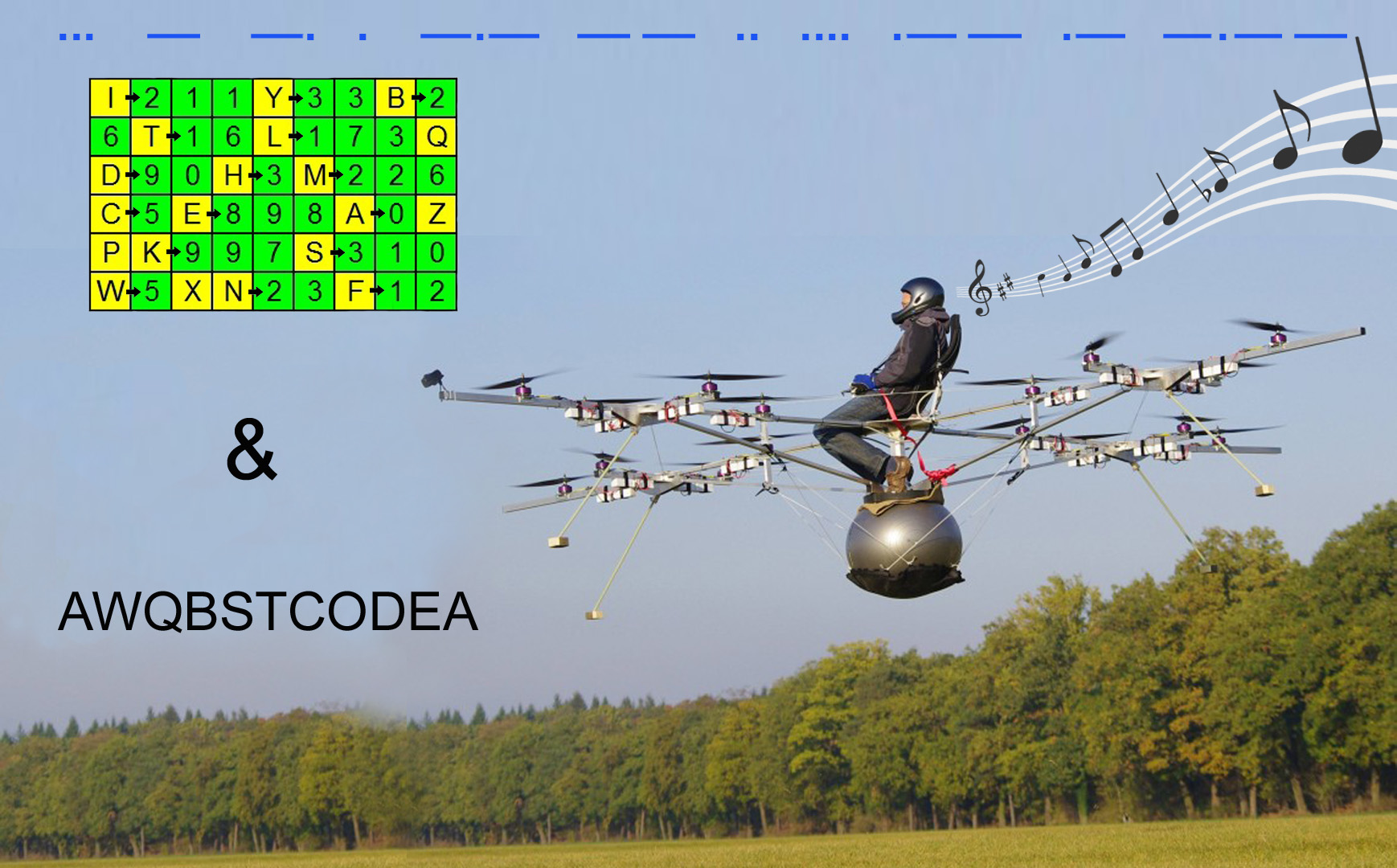}
\vspace{-7mm}
\caption{{\small Illustration for the problem ``A music lover''}}
\label{music}
\end{figure}

As usual Alex listens to music on the way to university. He chooses
it applying one secret code to the second one in his mind (Figure\,\ref{music}). Could you
understand what music he is listening  to right now?

{\bf Remarks.}

1. You should invent a way how to apply one code to another.

2. Some arithmetic operations also can be used.

\subsubsection{Solution}  Let us look at Figure\,\ref{music}. The first natural step in solving the problem is to decode the Morse Code at the top of the picture: {\tt STNEKMIHWAY}. The decoded string consists of 11 letters as well as the string {\tt AWQBSTCODEA} under the table. It brings to the mind that we should apply somehow one string to another using the information from the table. Note that almost every letter in the table has an arrow pointing to a sequence of numbers. And among the letters having arrows there are all the letters of the decoded string {\tt STNEKMIHWAY}:
\renewcommand{\arraystretch}{1.0}
\renewcommand{\tabcolsep}{1.0mm}
\begin{center}
\begin{tabular}{*{11}{l}}
{\tt S} $\rightarrow$ 3 1 0 & ~~~ &
{\tt T} $\rightarrow$ 1 6 & ~~~ &
{\tt N} $\rightarrow$ 2 3 & ~~~ &
{\tt E} $\rightarrow$ 8 9 8 & ~~~ &
{\tt K} $\rightarrow$ 9 9 7 & ~~~ &
{\tt M} $\rightarrow$ 2 2 6  \\
{\tt I} $\rightarrow$ 2 1 1 & ~~~ &
{\tt H} $\rightarrow$ 3 & ~~~ &
{\tt W} $\rightarrow$ 5 & ~~~ &
{\tt A} $\rightarrow$ 0 & ~~~ &
{\tt Y} $\rightarrow$ 3 3 & ~~~ &\\
\end{tabular}
\end{center}
Let us sum the numbers for each letter above! That is to calculate $3+1+0$, $1+6$, and so on. Thus, we get the string of eleven integers: $4, 7, 5, 25, 25, 10, 4, 3, 5, 0, 6$. Finally, we can apply this string to the {\tt AWQBSTCODEA} in the following way: each letter is cyclically shifting right in the alphabet by the corresponding number of positions from the integer string. We get
\begin{center}
\begin{tabular}{*{12}{r}}
letter & {\tt A}  & {\tt W}  & {\tt Q}  & {\tt B} & {\tt S} & {\tt T} & {\tt C} & {\tt O}  & {\tt D}  & {\tt E}  & {\tt A} \\
shift & 4 & 7 & 5 & 25 & 25 & 10 & 4 & 3 & 5 & 0 & 6 \\
result & {\tt E}  & {\tt D}  & {\tt V}  & {\tt A} & {\tt R} & {\tt D} & {\tt G} & {\tt R}  & {\tt I}  & {\tt E}  & {\tt G} \\
\end{tabular}
\end{center}
Thus, we conclude that Alex listens to the music of a great Norwegian composer Edvard Grieg.

The problem appeared to be difficult for the first round: it was completely solved by three participants. Nevertheless, twelve teams solved the problem during the second round.

\subsection{Problem ``The shortest addition chain''}

\subsubsection{Formulation}\label{adchain-1}

In many cryptographic systems we need to calculate the value
$B = A^{c} \textnormal{ mod } p,$
where $A$ is an integer, $1\leqslant A\leqslant p-1$, $c$ is an
arbitrary positive integer, and $p$ is a large prime number.
One possible way of reducing the computational load of calculating
is to minimize the total number of multiplications required to compute
the exponentiation. Since the exponent in the equation is additive, the problem of computing
powers of the base element $A$ can also be formulated as an addition
calculation, for which addition chains are used.

 An {\bf addition chain} for an integer $n$ is
a sequence of positive integers $a_0 = 1,$ $a_1,$ $\ldots,$ $a_{r-1},$ $a_r = n,$ where $r$ is a positive integer (that is called the {\bf length} of the addition chain) and  the following relation holds for all $i$, $1\leqslant i\leqslant r$:
$
a_i = a_j + a_k \text{ for some } k,j \text{ such that } k \leqslant j < i.
$

Find an addition chain of length as small as possible for the value $81$, present it as a list of values and mathematically prove that it can not be shorter!

\medskip

{\bf An example.} For the value 15 the shortest additional chain has length 5 and its list of values is $1, 2, 3, 6, 12, 15.$ So, to optimally calculate $B = A^{15} \textnormal{ mod } p,$ one can use just five multiplications:
$A^2=A\cdot A\!\!\mod{p}$, $A^3=A^2\cdot A\!\!\mod{p},$ $A^6=A^3\cdot A^3\!\!\mod{p},$  $A^{12}=A^6\cdot A^6\!\!\mod{p},$  $A^{15}=A^{12}\cdot A^3\!\!\mod{p}$.

\subsubsection{Solution}
It is easy to construct an addition chain of length 8 for 81, for example: $$1,2,4,8,16,32,64,80,81.$$
There are different ways of proving that a chain of length 7 or shorter does not exist. One way is making a computer program which will construct all possible addition chains of length $1, ..., 7$ and showing that none of them contains 81. We will provide a theoretical proof.

Let us have an addition chain $1=a_0,a_1,a_2,\ldots,a_r$ of length $r$. A trivial observation is that $a_k$ cannot be greater than $2^k$ for any $k$.
Thus, an addition chain of length $6$ or shorter cannot exist for number $81$, as $81 > 64 = 2^6$. So, we have to prove that an addition chain of length $7$ is not possible either. In order to do that, we prove the following lemma.

\noindent {\bf Lemma.}
    Let $1=a_0,a_1,a_2,\ldots,a_r$ be an addition chain of length $r$. Assume that $2^{r-1}<a_r\leqslant 2^r$. Then $a_r = 2^{r-1} + 2^s$ for some $0\leqslant s \leqslant r-1$.

\noindent{\bf Proof.}
    Let us prove the lemma by induction on $r$. For $r=1$, there is only one addition chain $1,2$, which satisfies the condition of the lemma.
    Assume that for all addition chains $a_0,a_1,a_2,\ldots,a_t$ of length $t<r$, such that $2^{t-1}<a_t\leqslant 2^t$ it holds $a_t = 2^{t-1} + 2^s$, where $0\leqslant s \leqslant t-1$.

    Let $1=a_0,a_1,\ldots,a_{t+1}$ be an addition chain of length $t+1$ such that $2^{t}<a_{t+1}\leqslant 2^{t+1}$. By definition of an addition chain, $a_{t+1}=a_n+a_m$ for some indices $n,m \leqslant t$. If both $a_n,a_m$ are not greater than $2^{t-1}$, then $a_{t+1}\leqslant 2^{t-1}+2^{t-1}=2^t$, that is a contradiction.

    Therefore, without loss of generality, $a_n > 2^{t-1}$. But that also means $n > t-1$. So, $n=t$ and the chain $a_0,a_1,\ldots,a_t$ satisfies the induction hypothesis. Therefore, $a_n = a_t = 2^{t-1} + 2^s$ for some $s\leqslant t-1$. Substituting $a_n$ into the expression for $a_{t+1}$, we obtain
    $$a_{t+1}=2^{t-1}+2^s+a_m$$
    for some $m\leqslant t$ and $s\leqslant t-1$. Now consider several cases:
    \begin{itemize}[noitemsep]
        \item{\underline{$s=t-1$}. In this case, $a_n=a_t=2^t$, which forces all $a_i$ to be equal to $2^i$. So, $a_{t+1}=2^t+2^m$ and the induction step is proven.}
        \item{\underline{$m=t$}. In this case, $a_{t+1} = 2a_t = 2^t + 2^{s+1}$ and the induction step is proven.}
        \item{\underline{$s<t-1$, $m=t-1$, $a_m \leqslant 2^{t-2}$}. In this case, $a_{t+1}=2^{t-1}+2^s+a_m \leqslant 2^{t-1}+2^{t-2}+2^{t-2}=2^t$, that is a contradiction.}
        \item{\underline{$s<t-1$, $m=t-1$, $a_m > 2^{t-2}$}. Then $a_0,a_1,\ldots,a_{t-1}$ satisfies the induction hypothesis, and $a_m = 2^{t-2} + 2^q$, $q\leqslant t-2$. Thus, $a_{t+1}=2^{t-1}+2^s+2^{t-2}+2^q$.

            --- If $q=t-2$, then $a_{t+1}=2^t+2^s$, and the induction step is proven.

            --- If $s=t-2$, then $a_{t+1}=2^t+2^q$, and the induction step is proven.

            --- If $q<t-2$ and $s<t-2$, then $a^{t+1}\leqslant 2^t$, which is a contradiction.}
        \item{\underline{$s<t-1$, $m<t-1$}. Then $a_m \leqslant 2^{t-2}$, and $a_{t+1}\leqslant 2^{t-1}+2^{t-2}+2^{t-2} = 2^t$, which is a contradiction.}
    \end{itemize}
    We considered all cases and checked that some of them contradict the lemma assumption, while others lead to the proven induction step. Thus, the lemma holds.\qed

    Now assume that there is an addition chain of length 7 for 81. Because $81>64 = 2^6$, such a chain would satisfy the condition of the lemma, and then $81=a_7=64+2^s$ for some $s\leqslant 6$. This is impossible. So, the shortest addition chain for 81 has length 8.

    This is a rather general solution, which also provides an interesting fact about addition chains.
    Just four school students completely solved the problem, the best solution was provided by  Alexander Dorokhin (Presidential PML 239, St. Petersburg, Russia). Usually, proofs of impossibility of a chain of length 7 were given in a more straightforward manner, checking possible strategies of getting to number 81 in 7 steps and showing that no matter how we add numbers, we will not be able to get 81.

\subsection{Problem ``An infinite set of collisions''}

\subsubsection{Formulation}

Bob is very interested in blockchain technology, so he decided to create his own system. He started with the construction of a hash function. His first idea for a hash function was the function $H$ with a hash value of length 16.

It works as follows.
\begin{itemize}[noitemsep]
    \item Let $u_1, u_2, \ldots, u_n \in \mathbb{F}_{2}$ be a data representation, $n$ is arbitrary.
    \item Bob calculates $z^0, \ldots, z^n \in \mathbb{F}_{2}^{32}$,
    $z^0 = (0, \ldots, 0)$, and $z^{i + 1}$ is obtained from $z^i$ in the following way:
\begin{center}
    $
        z' =
        \begin{cases}
(z^i_{1}, z^i_{2},\ldots, z^i_{16}, z^i_1 \oplus z^i_{17}, z^i_2 \oplus z^i_{18} \ldots, z^i_{16} \oplus z^i_{32}) & \text{ if } u_i = 1,\\

            (z^i_1 \oplus z^i_{17}, z^i_2 \oplus z^i_{18}, \ldots, z^i_{16} \oplus z^i_{32}, z^i_{17}, z^i_{18}, \ldots, z^i_{32}) & \text{ if } u_i = 0,\\
\end{cases}
    $

\smallskip
    $
        z'' =
        \begin{cases}
z' & \text{ if } u_i \neq z'_{32},\\

            (z'_1 \oplus 1, z'_2 \oplus 1, \ldots, z'_{32} \oplus 1) & \text{ if } u_i = z'_{32},\\
\end{cases}
    $
\smallskip

$
    z^{i + 1} = (z''_2, z''_3, \ldots, z''_{32}, u_i).
$
\end{center}
    \item Finally, $H(u_1, \ldots, u_n) = (z^{n}_1 \oplus z^{n}_{17}, z^{n}_2 \oplus z^{n}_{18}, \ldots, z^{n}_{16} \oplus z^{n}_{32})$.
\end{itemize}

But then Bob found out that his hash function is weak for using in crypto\-graphic applications.
Prove that Bob was right by constructing an infinite set $C \subset \bigcup_{n = 1}^{\infty} {\mathbb{F}_{2}^{n}}$ such that all elements of $C$ have the same hash value $H$.

\medskip {\bf An example.} Let us calculate $H(0, 1, 0)$. We have
\smallskip

$
    z^{1} = (\underbrace{1, 1, \ldots, 1}_{31}, 0),\ \ \ \
    z^{2} = (\underbrace{0, \ldots, 0}_{15}, \underbrace{1, \ldots, 1}_{15}, 0, 1),\ \ \ \
    z^{3} = (\underbrace{1, \ldots, 1}_{13}, 0, 0, \underbrace{1, \ldots, 1}_{14}, 0, 1, 0).
$
\smallskip

Thus, $H(0, 1, 0) = (\underbrace{0, \ldots, 0}_{14}, 1, 1).$

\subsubsection{Solution} Here we provide the solution proposed by Alexey Udovenko (University of Luxemburg). It consists of a theoretical proof and a simple example. Exactly the same idea was suggested by the program committee.

{\bf A theoretical proof}. Let us consider an arbitrary infinite sequence $u = u_1, u_2, u_3, \ldots$ and the following hash values:
$$H(u_1), H(u_1, u_2), H(u_1, u_2, u_3), \ldots$$
To obtain an infinite set of collisions, it is enough to find some $\ell, m$, such that $z^{\ell} = z^{m}$. Then we reconstruct the sequence in the following way: $u_{k} = u_{k - |m - \ell|}$ starting with $k = \max\{\ell, m\} + 1$.
By this method we obtain a cycle in the sequence of the states  $z^1, z^2, z^3, \ldots$, since each state $z^i$ uniquely defines $H(u_1, \ldots, u_i)$ and the next state $z^{i + 1}$ in conjunction with $u_{i+1}$. The cycle length divides $2^{32}!$, since it is between $1$ and $|\mathbb{F}_2^{32}| = 2^{32}$. The initial state $z_0$ may not belong to the cycle, but after $2^{32}$ steps $z^{2^{32}}$ definitely belongs to the cycle. It means that
$$
    H(u_1, \ldots, u_{2^{32}}) = H(u_1, \ldots, u_{2^{32} + 1 \cdot 2^{32}!}) = H(u_1, \ldots, u_{2^{32} + 2 \cdot 2^{32}!}) = \ldots
$$

{\bf
An example.} Let us consider the zero sequence $u = 0, 0, 0, \ldots$. In this case
$$
    z^{31} = (0, 1, 1, 0, 0, 1, 1, 0, 0, 1, 1, 0, 0, 1, 1, 0, 1, 0, 1, 0, 1, 0, 1, 0, 1, 0, 1, 0, 1, 0, 1, 0)
$$
and $z^{32} = z^{31}$. Thus, $H(u_1, \ldots, u_k) = (1, 1, 0, 0, 1, 1, 0, 0, 1, 1, 0, 0, 1, 1, 0, 0)$ for any $k \geqslant 31$.

The problem was completely solved by twelve university students and professionals.

\subsection{Problem ``One more parameter''}

\subsubsection{Formulation}

There are several parameters in cryptanalysis of block ciphers that are used to measure the diffusion strength. In this problem, we study properties of one of them.

Let $n$, $m$ be positive integers. Let $a = (a_1, \dots , a_m)$ be a vector with coordinates
$a_i$ taken from the finite field $\mathbb{F}_2$.
Denote the number of nonzero coordinates
$a_i$, $i = 1,\dots,m$, by $\mathrm{wt}(a)$ and call this number the {\bf weight} of
the vector $a$. The inner product of
$a=(a_1,\dots,a_m)$ and $b=(b_1,\dots,b_m)$ in $\mathbb{F}^m_{2}$
 is defined as
$a \cdot b =  a_1b_1 \oplus \ldots \oplus a_mb_m.$
For a Boolean function $f: \mathbb{F}^m_{2} \rightarrow \mathbb{F}_{2}$,
we define the function {\bf weight}, $\mathrm{wt}$, as follows:
$
\mathrm{wt}(f) =  |\{a \in  \mathbb{F}^m_{2}\ |\   f(a) = 1\}|.
$

The {\bf special parameter} $Q$ of a vectorial Boolean function
$\varphi: \mathbb{F}^m_{2} \rightarrow \mathbb{F}^m_{2}$ is defined to be
\[
Q (\varphi) = \min\limits_{a,\,b,\, b \neq {\bf 0}, \,  \mathrm{wt}(a \cdot x \oplus b \cdot \varphi(x)) \neq 2^{m-1}}
\{\mathrm{wt}(a) + \mathrm{wt}(b)\}.
\]
\begin{itemize}[noitemsep]
\item Rewrite (simplify) the definition of $Q(\varphi)$ when the function
$\varphi$ is linear (recall that a function $\ell$ is linear if
$\ell(x \oplus y) = \ell(x) \oplus \ell(y)$ for any $x, y$).
\item Rewrite the definition of $Q(\varphi)$ in terms of linear codes, when
 the linear function $\varphi$ is given by an $m \times m$ matrix $M$
 over $\mathbb{F}_2$, i.\,e. $\varphi(x)=M x$.
\item Find the tight upper bound for $Q(\varphi)$ as a function of $m$.
\item Can you give an example of the function $\varphi$ with the
maximal possible value of $Q$?
\end{itemize}

\subsubsection{Solution} A special parameter $Q$ considered in the problem is called the {\it linear branch
number} of a transformation \cite{aes}. This problem is a linear cryptanalysis equivalent of the problem ``A special parameter'' of NSUCRYPTO'2014 \cite{paper-pdm}, where the differential branch number was discussed.

Let $\varphi$ be a vectorial Boolean function $\mathbb{F}_2^m\rightarrow\mathbb{F}_2^m$.

$\bullet$ If $\varphi$ is a linear function, then  the Boolean function $a\cdot x\oplus b\cdot\varphi(x)$ is also linear for any vectors $a,b\in\mathbb{F}_2^m$. Hence, the condition
$\text{wt}(a\cdot x\oplus b\cdot\varphi(x))\ne 2^{m-1}$
is equivalent to
$   a\cdot x\oplus b\cdot\varphi(x)=0$ for all $x\in\mathbb{F}_2^m$.
    Thus, for the considered case we have the definition
    $$
    Q(\varphi)=\min\limits_{a,\,b,\,b\ne {\bf 0},\,a\cdot x\oplus b\cdot\varphi(x)\equiv0}\{\text{wt}(a)+\text{wt}(b)\}.
$$

$\bullet$ Let us consider vectors as columns. In the case when $\varphi(x)=Mx$ for some $m\times m$ matrix $M$ over the field $\mathbb{F}_2$ we can rewrite $a\cdot x\oplus b\cdot\varphi(x) = (a \oplus M^{T}b)\cdot x$. Then, $(a \oplus M^{T}b)\cdot x\equiv 0$ implies $a \oplus M^{T}b = {\bf 0}$ or $Hc = {\bf 0}$, where $H=(I|M^{T})$ is a $m\times 2m$ matrix, $I$ is the identity $m\times m$ matrix, and $c = (a,b)$ denotes the concatenation of vectors $a$ and $b$ of length $2m$. Note, that $b={\bf 0}$ and $a = M^{T}b$ imply $a={\bf 0}.$ So, $b\neq{\bf 0}$ is equivalent to $c\neq{\bf 0}$. Thus,
$$
    Q(\varphi)=\min\limits_{c\neq {\bf 0},\, Hc = {\bf 0}}\{\text{wt}(c)\} = {\rm dist}(C),
$$
where $C$ is the linear code of length $2m$ and dimension $m$ with a parity-check matrix $H$, ${\rm dist}(C)$ denotes the distance of the code $C$.

$\bullet$ Here we would like to apologize to the participants since the formulation of the problem was not correctly stated. If we consider a mapping $\varphi:\mathbb{F}_{2^n}^m\to\mathbb{F}_{2^n}^m$ instead of $\varphi:\mathbb{F}_{2}^m\to\mathbb{F}_{2}^m$, then one can easily find the bound $Q(\varphi)\leqslant m+1$, and this bound is tight for various parameters $n$ and $m$. Indeed, there exist Maximal Distance Separable codes with parameters $[2m, m, m + 1]$ over $\mathbb{F}_{2^n}$ (for example, Reed--Solomon codes).

At the same time, for $\varphi:\mathbb{F}_{2}^m\to\mathbb{F}_{2}^m$ as it was given in the problem, the bound $Q(\varphi) \leqslant m+1$ can be achieved only when $m=1$. So, we cannot say that this bound is tight for various $m$. So, this bound cannot be considered as a correct answer.

To be honest, we could not say what is the correct answer to this problem, and so, we may assume that this problem is also one of the open problems of the Olympiad. What was surprising and very pleasant for us is that several teams found nontrivial bounds for general and linear cases. But unfortunately, they could not say if these bound are tight.
We would like to shortly present the main results of the participants.

{\bf The linear case:} $Q(\varphi)\leqslant(2m+4)/3$.
This bound was found by Irina Slonkina (National Research Nuclear University MEPhI).

    Let us consider any $m\times m$ matrix $S$ over the field $\mathbb{F}_2$ and a linear function $\varphi_s(x)=Sx$. So,
$$
    Q(\varphi_s)=\min\limits_{b\ne {\bf 0}}\{\text{wt}(bS)+\text{wt}(b)\}.
$$
    It is clear that for any $i,j\in\{1,2,...,m\}$, $i\neq j$, the following bounds hold:
$$
    Q(\varphi_s)\leqslant\text{wt}(S_i)+1\ \
\text{ and }\ \
    Q(\varphi_s)\leqslant\text{wt}(S_i\oplus S_j)+2,
$$
    where $S_i,S_j\in\mathbb{F}_2^m$ are $i$-th and $j$-th rows of the matrix $S$. Then it holds
    \begin{multline*}
    Q(\varphi_s)-2\leqslant\text{wt}(S_i\oplus S_j)\leqslant 2m-\text{wt}(S_i)-\text{wt}(S_j)\leqslant\\\leqslant 2m-(Q(\varphi_s)-1)-(Q(\varphi_s)-1)
    =2m-2Q(\varphi_s)+2.
    \end{multline*}
    where the second bound follows from the inequality $\text{wt}(u\oplus v)\leqslant 2m-\text{wt}(u)-\text{wt}(v)$ that holds for any $u,v\in\mathbb{F}_2^m$.
    Thus, for the function $\varphi_s$ we have the bound
$   Q(\varphi_s)\leqslant(2m+4)/3.$

\medskip
    {\bf The general case:} if $m\geqslant 2$, then $Q\left(\varphi\right)\leqslant m$.

 This bound was found by Alexey Miloserdov, Saveliy Skresanov, and Nikita Odinokih team (Novosibirsk State University) and by Kristina Geut and Sergey Titov team (Ural State University of Railway Transport). Here, we present the solution of the first team.

        Let $f:\mathbb{F}_2^n\to\mathbb{F}_2$. The {\it Walsh transform} of $f$ is defined as $W_f(y)=\sum\limits_{x\in\mathbb{F}_2^n}(-1)^{y\cdot x\oplus f(x)},\ y\in\mathbb{F}_2^n.$
    The function $f$ is uniquely defined by its {\it Walsh coefficients} since the following equality holds:
$$
        (-1)^{f(x)}=\frac{1}{2^n}\sum\limits_{y\in\mathbb{F}_2^n}W_f(y)(-1)^{y\cdot x},\ x\in\mathbb{F}_2^n.
$$
    It is also well known that {\it Parseval's equality},
$\sum\limits_{y\in\mathbb{F}_2^n}W_f^2(y)=2^{2n}$, holds for any Boolean function~$f$.

\smallskip
\noindent{\bf Proposition.} Let $f:\mathbb{F}_2^n\to\mathbb{F}_2$. Suppose that for every $a\in\mathbb{F}_2^n$ such that $0\leqslant{\rm wt}(a)\leqslant n-1$ it holds that ${\rm wt}(a\cdot x\oplus f(x))=2^{n-1}$. Then $f(x)=x_1\oplus x_2\oplus\ldots\oplus x_n\oplus c$ for some $c\in\mathbb{F}_2$.
\smallskip

\noindent{\it Proof.} Since a function $a\cdot x\oplus f(x)$ is balanced iff $W_f(a)=0$, then by Parseval's equality and the assumption of the proposition we have $W_f(a)=0$ for all $a\in\mathbb{F}_2^n$ such that $0\leqslant{\rm wt}(a)\leqslant n-1$ and $|W_f({\bf 1})|=2^n$, where ${\bf 1}=(1,1,\ldots,1)\in\mathbb{F}_2^n$. In this case it holds
$
        (-1)^{f(x)}=\frac{\pm 2^n}{2^n}(-1)^{{\bf 1}\cdot x},\ x\in\mathbb{F}_2^n,
$
        i.\,e.
$
        f(x)=x_1\oplus x_2\oplus\ldots\oplus x_n\oplus c,
$
        for some $c\in\mathbb{F}_2$.\qed

    \medskip
\noindent{\bf Corollary.} $Q_{\max}(m)\leqslant m$ for $m\geqslant 2$, where $Q_{\max}(m)=\max\limits_{\varphi:\mathbb{F}_2^m\rightarrow\mathbb{F}_2^m}Q(\varphi)$.
    \smallskip

    \noindent{\it Proof.} Denote $b_1=(1,0,0,\ldots,0)\in\mathbb{F}_2^m$, $b_2=(0,1,0,\ldots,0)\in\mathbb{F}_2^m$.
        Assume that there exists some $\varphi$ such that $Q(\varphi)>m$. It implies $\text{wt}(a\cdot x\oplus b_i\cdot\varphi(x))=2^{m-1}$, $i=1,2$, for any constant $a\in\mathbb{F}_2^m$ such that $0\leqslant\text{wt}(a)\leqslant m-1$. Then by the Proposition it holds
$
        b_i\cdot\varphi(x)={\bf 1}\cdot x\oplus c_i,
$
        for some $c_i\in\mathbb{F}_2$, $i=1,2$. Thus, the sum modulo $2$ of the first and the second coordinate functions of $\varphi$ is a constant function. Hence,
$
        \text{wt}(b\cdot\varphi(x))\in\{0,2^m\},
$
        where $b=(1,1,0,\ldots,0)\in\mathbb{F}_2^m$. But then we have $Q(\varphi)\leqslant 2$, that is a contradiction.\qed

\subsection{Problem ``Scientists''}

\subsubsection{Formulation}

Two young cryptographers and very curious students Alice and Bob studied different cryptosystems and  attacks on them. At the same time they were very interested in biographies of famous scientists and found out one interesting {\bf property} that can be used in cryptosystems.
They choose three pairs of scientists:
\begin{center}
\renewcommand\tabcolsep{1mm}
\begin{tabular}{rcl}
{\bf Charles Darwin} & and & {\bf Michael Faraday},\\

{\bf Werner Heisenberg} & and & {\bf Johannes Kepler},\\

{\bf Hans Christian Orsted} & and & {\bf Mikhail Lomonosov}.\\
\end{tabular}
\end{center}

Alice and Bob choose a cryptosystem and an attack they would like to study. They constructed three sets of parameters for the cryptosystem: one set according to each pair of scientists. Then Alice chooses a phrase consisting of 18 English letters (spaces were omitted) and divided it into three parts of 6 letters. She represented each part as a hexadecimal number using ASCII code. Alice encrypted the first part by the cryptosystem for each set of parameters, then the same actions she made for the second and the third parts. Finally, Alice got the following three groups of three ciphertexts (in hexadecimal notation):
\begin{center}
\begin{tabular}{cccc}
& Part 1 & Part 2 & Part 3 \\
Set of parameters 1 & {\tt 2512 1F5A 0079} & {\tt B494 222D 3E1C} & {\tt 275E B751 4FDB} \\

Set of parameters 2 & {\tt 3D0D 6812 0443} & {\tt 5111 5BFD 9398} & {\tt 0815 6223 2698} \\

Set of parameters 3 & {\tt 1EDC 4856 8CE2} & {\tt 9C18 2A32 B9AB} & {\tt 9A1C AD5C 25D7} \\
\end{tabular}
\end{center}
and asked Bob to decrypt it using the attack! Bob successfully read the secret phrase. Could~you

\begin{itemize}[noitemsep]
\item find the property like Alice and Bob,

\item understand what is the cryptosystem and the attack chosen,

\item decrypt the ciphertext by applying this attack?
\end{itemize}

\medskip

 *What word should be added at the beginning of the decrypted text according to the famous words of Mikhail Lomonosov?

\subsubsection{Solution} The problem is related to the RSA cryptosystem~\cite{RSA1978} and the broadcast attack
due to Hastad~\cite{Hastad1988}. Given pairs of famous scientists, some participants made a correct guess
that each of the pairs is linked to some prime numbers $P$, $Q$, and the RSA modulus $N=PQ$.
But what is the way to obtain the prime numbers?! Success in their search depends only on one's
intuition. Writing down the birth date of each of the scientists in the form DDMMYYYY, one can
notice that all these 7- or 8-decimal numbers are prime:
\begin{center}
\begin{tabular}{lcl}
$P_1=12021809$ & and & $Q_1=22091791$,
\\
$P_2=5121901$ & and & $Q_2=27121571$,
\\
$P_3=14081777$ & and & $Q_3=19111711$.
\end{tabular}
\end{center}
This is the property Alice and Bob found out from biographies of the scientists.

Since Alice encrypted each part of the original text thrice, in order to decrypt it
we can try to apply Hastad's broadcast attack on RSA as it is described in~\cite{Boneh1999}.
Three pairs of parameters should indicate that $e=3$ was chosen as the public exponent.
This is not a good decision of Alice and Bob. A valid public exponent must be
coprime with $\varphi(N)$, since that makes it possible to compute the private exponent.
Whereas $3$ divides either of $\varphi(N_1)$, $\varphi(N_2)$, or $\varphi(N_3)$.
Two solutions noted this weird choice and both of them are marked as the best solutions.

Nevertheless, it is still reasonable to use Chinese Remainder Theorem and take the cubic root.
In other words, applying Hastad's broadcast attack we obtain three parts of the message encrypted.
Converting them back to ASCII characters we get the plaintext {\tt PUTSTHEMINDINORDER},
which means {\tt PUTS THE MIND IN ORDER}. This is a part of the famous phrase by a distinguished Russian scientist Mikhail Lomonosov, who said that ``Mathematics should be studied because it puts the mind in order''. Consequently, the first word of the quote is ``mathematics''.

The participants presented two comprehensive solutions at the first round and seven yet at the second round. One more solution turned to be almost complete: authors pointed a wrong word as the beginning of the Lomonosov statement. The best solutions were proposed by Daniel Malinowski and Michal Kowalczyk team (University of Warsaw, Dragon Sector) and by Alexey Ripinen, Oleg Smirnov, and Peter Razumovsky team (Saratov State University).

\subsection{Problem ``Masking''}

\subsubsection{Formulation}

It is known that there are attacks on cryptosystems that use information obtained from the physical implementation of a cryptosystem, for example, timing information, power consumption, electromagnetic leaks or even sound. To protect cryptosystems from such attacks cryptographers can use a countermeasure known as {\bf masking}.

Correlation immune Boolean functions can reduce the masking cost. Therefore, we need to search for Boolean functions satisfying the following conditions: they should have {\bf small Hamming weight}, for implementation reasons, and {\bf high correlation immunity} to resist an attacker with multiple probes.

Let $f$ be a non-constant Boolean function in 12 variables of correlation immunity equal to~6.
\begin{itemize}[noitemsep]
\item What is the lowest possible Hamming weight $k$ of $f$?
\item Give an example of such a function $f$ with Hamming weight $k$.
\end{itemize}

{\bf Remarks.}

1. Hamming weight ${\rm wt}(f)$ of a Boolean function $f$ in $n$ variables is the number of vectors $x\in\mathbb{F}_2^n$ such that $f(x) = 1$.

2. A Boolean function $f$ in $n$ variables is called {\it correlation immune of order $t$}, where $t$ is an integer such that $1\leqslant t \leqslant n$, if ${\rm wt}(f_{i_1,\ldots,i_t}^{a_1,\ldots,a_t})={\rm wt}(f)/2^t$ for any set of indexes  $1\leqslant i_1 < \ldots < i_t \leqslant n$ and any set of values $a_1,\ldots,a_t\in \mathbb{F}_2$. Here $f_{i_1,\ldots,i_t}^{a_1,\ldots,a_t}$ denotes the subfunction of $f$ in $n-t$ variables that is obtained from $f(x_1,\ldots,x_n)$ by fixing each variable $x_{i_k}$ by the value $a_k$, $1\leqslant k\leqslant t$.

\subsubsection{Solution} Firstly, we should note that this problem contains open questions in general. We considered the participants' solutions as correct if they are as deep as solutions known to the Olympiad program committee. More precisely, we expected from the participants Boolean functions in 12 variables of weight 1024 that are correlation immune of order 6.

 Let $f$ be a non-constant Boolean function in $n$ variables of correlation immunity~$t$ and of Hamming weight $k$. The known open problem is to find such a function $f$ having as low as possible Hamming weight for various $n$ and $t$. The problem questions were investigated in \cite{masking-1, masking-2}, where minimal Hamming weight $k=1024$ of $f$ for $n=12$ and $t=6$  was found using heuristics (more precisely, evolutionary algorithms). Theoretically, $k$ can be lower than 1\,024 but it is unknown any example of such a function. It can be equal to any value of the form $k=64\ell$ greater than or equal to  768 according to the results on orthogonal arrays \cite{masking-3}. It is known that the elements of the support of $f$ form the rows of an orthogonal array with parameters $(k,n,2,t)$ (recall that $x\in\mathbb{F}_2^n$ belongs to the support of $f$ if $f(x)=1$).

We present several constructions of $f$ proposed by the participants.

1) The first compact example was obtained by Maxim Plushkin, Ivan Lozinskiy, and Azamat Miftakhov team (Lomonosov Moscow State University). They found the following function:
  $$f(x_1,\ldots,x_{12}) = (x_1 \oplus x_2 \oplus x_3 \oplus x_4 \oplus x_5 \oplus x_6 \oplus x_7 \oplus 1) (x_6 \oplus x_7 \oplus x_8 \oplus x_9 \oplus x_{10} \oplus x_{11} \oplus x_{12} \oplus 1).$$

  Note that the team studied the problem for a small number of variables $n$ up to 14 and correlation immunity $t=n/2$. For example, they found a function in $10$ variables with the Hamming weight $k=256$ constructed similarly to the case of 12 variables: $$f(x_1,\ldots,x_{10}) = (x_1 \oplus x_2 \oplus x_3 \oplus x_4 \oplus x_5 \oplus x_6 \oplus 1) (x_5 \oplus x_6 \oplus x_7 \oplus x_8 \oplus x_9 \oplus x_{10} \oplus 1).$$ And as proved in \cite{masking-1} this weight cannot be lower.

2) Another solution was found by Alexey Udovenko (University of Luxembourg) in the following way:
$$f(x_1,\ldots, x_{12}) = s_1 \oplus s_1s_2 \oplus s_1s_3 \oplus s_2s_3,$$
where
$
   s_1 = x_1 \oplus x_2 \oplus x_3 \oplus x_4,\
   s_2 = x_5 \oplus x_6 \oplus x_7 \oplus x_8,\
   s_3 = x_9 \oplus x_{10} \oplus x_{11} \oplus x_{12}.
$

Alexey also mentioned that in the case of quadratic function in 12 variables the Hamming weight cannot be less than 1024 (Proposition\,1.9 \cite{masking-4}). Note that he concentrated his search on quadratic functions whose graphs of quadratic terms have multiple automorphisms. This idea was supported by studying the graph of quadratic terms of a function $f$ with parameters $n=6$, $t=3$ and $k=16$. Alexey computationally proved that in this case $16$ is the minimal Hamming weight of $f$.

3) The third interesting example was proposed by Anna Taranenko (Sobolev Institute of Mathematics) and can be described as follows:
$$f(x_1,\ldots,x_{12}) = 1 \Leftrightarrow
\varphi(x_1,x_2,x_3) + \varphi(x_4,x_5,x_6) + \varphi(x_7,x_8,x_9) + \varphi(x_{10},x_{11},x_{12}) = 0,$$
where $+$ denotes the addition in $\mathbb{Z}_4$ and $\varphi$ takes the following values:
\begin{center}
$\varphi(0,0,0)=\varphi(1,1,1) = 0;\ \varphi(1,0,0)=\varphi(0,1,1) = 1;$
$\varphi(0,1,0)=\varphi(1,0,1) = 2;\ \varphi(0,0,1)=\varphi(1,1,0) = 3.$
\end{center}
Anna presented a mathematical proof that the function $f$ is correlation immune of order $7$ (and therefore $6$) with the Hamming weight 1024. She also mentioned that for a 12-variables function of correlation immunity $7$ the minimal Hamming weight is exactly 1024 according to the Bierbrauer~---~Friedman inequality for parameters of orthogonal arrays \cite{masking-5, masking-6}.

Over all, only these three teams mentioned above made significant progress with this problem.

\subsection{Problem ``{\tt TwinPeaks}''}

\subsubsection{Formulation}

On Bob's smartphone there is a program that encrypts messages with the algorithm \texttt{TwinPeaks}. It works as follows:
\begin{enumerate}[noitemsep]
\item[{\bf 1.}] It takes an input message $P$ that is a hexadecimal string of length~$32$ and represents it as a binary word $X$ of length~$128$.

\item[{\bf 2.}] Then $X$ is divided into four $32$-bits words $a,b,c,d$.

 \item[{\bf 3.}] Then six rounds of the following transformation are applied:
$$
(a,b,c,d)\leftarrow
\big(
a+ c+ S(c+ d),
a+ b+ d+ S(c+ d),
a+ c+ d,
b+ d+ S(c+ d)
\big),
$$
where~$S$~is a secret permutation from $\mathbb{F}_2^{32}$ to itself and
 $+$ denotes the coordinate-wise sum modulo 2.

\item[{\bf 4.}] The word $Y$ is obtained as a concatenation of $a,b,c,d$.

\item[{\bf 5.}] Finally, $Y$ is converted to the hexadecimal string $C$ of length $32$. The algorithm gives $C$ as the ciphertext for $P$.
\end{enumerate}

Agent Cooper intercepted the ciphertext
$C=\texttt{59A0D027D032B394A0A47A9ED19C98A8}$ sent from Bob to Alice
and decided to decrypt it.

In order to solve this problem agent Cooper also captured Bob's
smartphone with the \texttt{TwinPeaks} algorithm!
\href{https://nsucrypto.nsu.ru/archive/2017/round/2/task/2}{\textcolor{blue}{Here}}
it is. Now Cooper (and you too) can encrypt any messages with
\texttt{TwinPeaks} but still can not decrypt any one.

Help Cooper to decrypt $C$.

\subsubsection{Solution}
Let $F$ be the round transformation of \texttt{TwinPeaks}:
$$
F(a,b,c,d)=
(
a+ c+ S(c+ d),
a+ b+ d+ S(c+ d),
a+ c+ d,
b+ d+ S(c+ d)
)
$$
and
$$
f(a,b,c,d)=(c+ d,a+ b+ c,a+ b,b+ c+ d).
$$
If $F$ transforms a message~$(a,b,c,d)$ to~$(a',b',c',d')$,
then $G$ tranforms a message~$f(a,b,c,d)$ to~$f(a',b',c',d')$, where $G$ acts as follows:
$$
G(a,b,c,d)=(b+S(a),c,d,a).
$$
This conclusion can be extended to all six rounds.
It will be convenient to consider a modification of~\texttt{TwinPeaks},
where~$F$ is replaced by~$G$. Indeed, one can encrypt a message~$(x_1,x_2,x_3,x_4)$, where
$x_i\in\mathbb{F}_2^{32}$, using the modified algorithm in the following way:
\begin{itemize}[noitemsep]
\item[{\bf 1.}] Encrypt
$f^{-1}(x_1,x_2,x_3,x_4)=(x_1+x_3+x_4,x_1+x_4,x_2+x_3,x_1+x_2+x_3)$
with~\texttt{TwinPeaks}.
\item[{\bf 2.}]
 Transform the encryption result using~$f$.
\end{itemize}

\begin{floatingfigure}[r]{5.2cm}
\vspace{-0.4cm}
\hspace{0.1cm}\includegraphics[width=4.2cm]{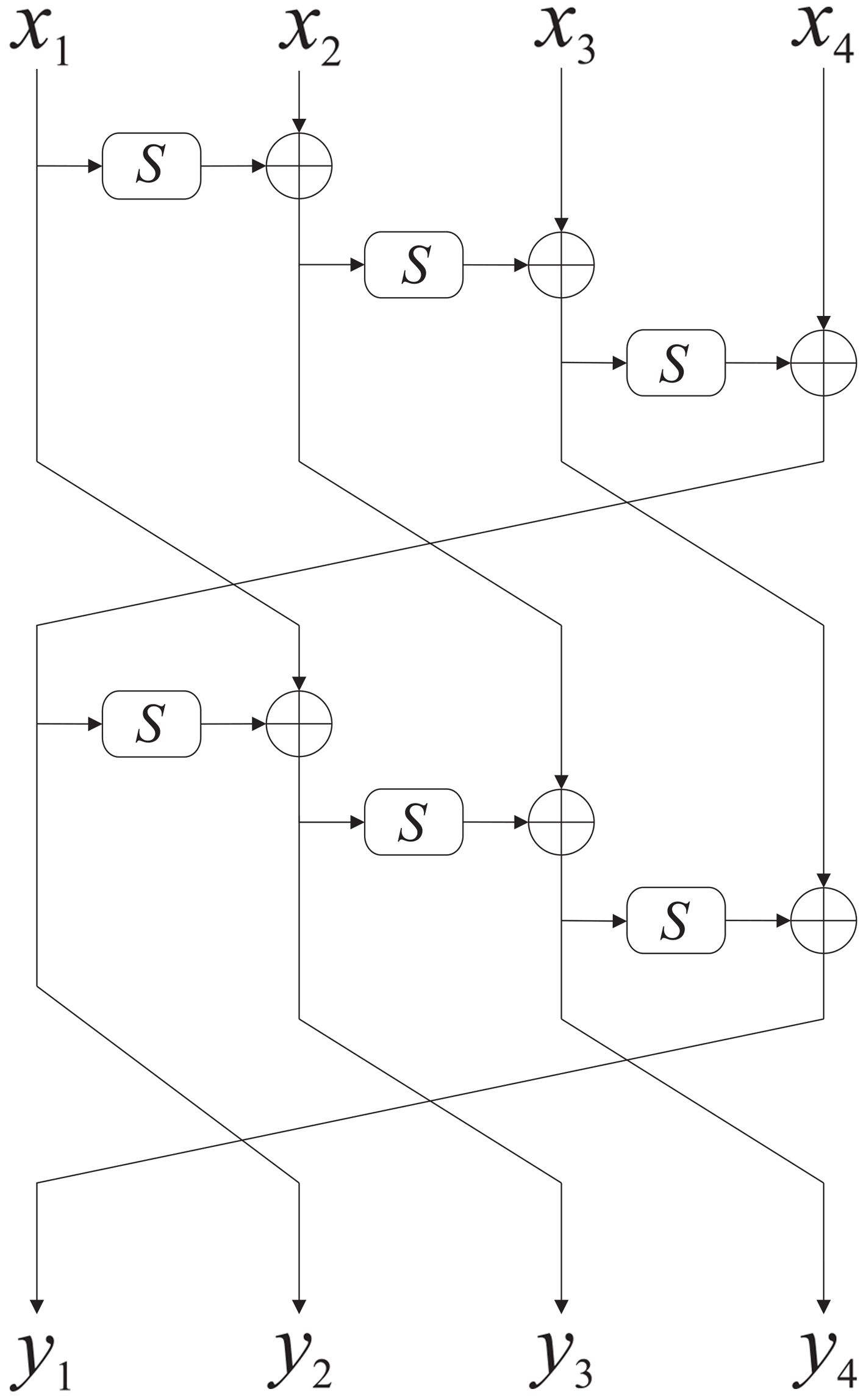}
\caption{{\small Modified~\texttt{TwinPeaks}}}
\label{twin}
\end{floatingfigure}
Figure\,\ref{twin} illustrates the modified~\texttt{TwinPeaks}.
As one can see, plaintexts and ciphertexts are linked with the following relations:
\begin{align*}
y_1&=x_3+S(x_2+S(x_1))+S(y_4),\\
y_2&=x_4+S(x_3+S(x_2+S(x_1))),\\
y_3&=x_1+S(y_2),\\
y_4&=x_2+S(x_1)+S(y_3).
\end{align*}

Cooper can choose as~$x_1$ any value~$u$ and get its representation in the form $u=v+S(w)$, where $v=y_3$, $w=y_2$.
Moreover, Cooper can represent $w$ in the form~$v'+S(w')$ and finally get a representation
$$
u=v+S(v'+S(w')).
$$
Suppose that Cooper wants to find~$x_1$. Then he represents
$$
y_2=v+S(v'+S(w'))
$$

\noindent(2~requests to \texttt{TwinPeaks}) and encrypts $(w',v',v,y_3)$
(1~request).
The second word of the ciphertext obtained is
$$
y_3+S(v+S(v'+S(w')))=y_3+S(y_2)=x_1.
$$

Since Cooper is able to find~$x_1$ given ~$(y_1,y_2,y_3,y_4)$, he can calculate~$S(u)$ for all~$u$.
Indeed, Cooper can choose~$y_2=u$ and
arbitrary~$y_1,y_3,y_4$. Then he finds~$x_1$ and~$S(u)=x_1+y_3$.
It can be done using 3 requests.

Using 6 requests Cooper can find~$S(x_1)+S(y_3)$ and, hence, $x_2$.
Another 6 requests is enough to find $S(x_2+S(x_1))+S(y_4)$ and, hence, $x_3$.
Finally, using 3 request he finds $S(x_3+S(x_2+S(x_1)))$ and, hence, $x_4$.

Thus, one needs $3+6+6+3=18$ requests to decrypt the message.

The answer is  {\tt 43ABECCAA53CB953F35239E79CC900EE}.

Correct solutions were proposed by twenty teams of university students and professionals. All of them used different methods and techniques. We did not identify a single best but we are pleased to note the participants professionalism and creativity in solving the problem.

\subsection{Problem ``An addition chain''}

\subsubsection{Formulation}

In many cryptographic systems we need to calculate the value
$B = A^{c} \textnormal{ mod } p,$
where $A$ is an integer, $1\leqslant A\leqslant p-1$, $c$ is an
arbitrary positive integer, and $p$ is a large prime number.
One possible way of reducing the computational load of calculating
is to minimize the total number of multiplications required to compute
the exponentiation. Since the exponent in equation is additive, the problem of computing
powers of the base element $A$ can also be formulated as an addition
calculation, for which addition chains are used.

 An {\bf addition chain} for an integer $n$ is
a sequence of positive integers $a_0 = 1,$ $a_1,$ $\ldots,$ $a_{r-1},$ $a_r = n,$ where $r$ is a positive integer (that is called the {\bf length} of the addition chain) and  the following relation holds for all $i$, $1\leqslant i\leqslant r$:
$
a_i = a_j + a_k \text{ for some } k,j \text{ such that } k \leqslant j < i.
$

Find an addition chain of length as small as
possible for the value $2^{127} - 3.$

The solution should be
submitted as a list of values occurring in the chain and a
description of how you found the solution.
An example of the shortest addition chain for the value 15 can be found in section \ref{adchain-1}.

\subsubsection{Solution} We should note that the problem contains open questions in general. We considered the participants' solutions as correct if they are as deep as solutions known to the Olympiad program committee. More precisely, if the participants could find an addition chain of length~136.

We would like to follow the solution proposed by Alexey Miloserdov, Saveliy Skresanov, and Nikita Odinokih team (Novosibirsk State University). Denote by $\ell(n)$ the length of the smallest addition chain for a number $n$. Let us first prove that $\ell(2^{127} - 3)\leqslant 136$ and present a chain of length 136. Then we will consider several lower bounds for $\ell(2^{127} - 3)$.

It is easy to see that $$2^{127} - 3 = 4 (2^{125} - 1) + 1.$$
We have two inequalities for any $n,m\geqslant11$: $\ell(n + 1) \leqslant \ell(n) + 1$ and $\ell(nm) \leqslant \ell(n) + \ell(m)$.
So, we can conclude that $\ell(2^{127} - 3) \leqslant \ell(2^{125} - 1) + 3$, since $\ell(4) = 2$.

An addition chain $a_0 = 1, a_1, \ldots, a_{r-1}, a_r = n$ is called a {\it star chain} for $n$ if for each $1\leqslant i \leqslant r$ there exists $0\leqslant j < i$ such that $a_i = a_{i-1} + a_j$. Denote by $\ell^{*}(n)$ the length of the shortest star chain for the number $n$. It is easy to see $\ell(n)\leqslant \ell^{*}(n)$. The following inequality holds by the famous Brauer's theorem~\cite{chain-1}: $\ell(2^m-1)\leqslant m-1+\ell^{*}(m)$ for any $m\geqslant1$.
The chain $1, 2, 3, 5, 10, 20, 25, 50, 100, 125$ is a star chain for $125$ of length $9$. It is known \cite{chain-2} that $\ell(125) = 9$. So, the chain for $125$ found above is the shortest one.
Thus, $$\ell(2^{127}-3)\leqslant 125 - 1 + 9 + 3 = 136.$$
A required chain of length 136 for $2^{127}-3$ is presented in Table~\ref{chain}.

\begin{table}[h]
\centering\tiny
\caption{An addition chain of length 136 for $2^{127}-3$.}
\smallskip
\label{chain}
\begin{tabular}{|p{15cm}|}
  \hline
  1, 2, 4, 8, 12, 24, 48, 60, 120, 124, 248, 496, 992, 1984, 3968, 4092, 8184, 16368, 32736, 65472, 130944, 261888,   523776, 1047552, \newline
    2095104, 4190208, 4194300, 8388600, 16777200, 33554400, 67108800, 134217600, 134217724, 268435448,   536870896, 1073741792, \newline
  2147483584, 4294967168, 8589934336, 17179868672, 34359737344, 68719474688, 137438949376,
   274877898752, 549755797504,\newline
   1099511595008, 2199023190016, 4398046380032, 8796092760064, 17592185520128, 35184371040256,
     70368742080512,  \newline
     140737484161024, 281474968322048, 562949936644096, 1125899873288192, 2251799746576384, 4503599493152768, 4503599627370492,\newline
      9007199254740984,  18014398509481968, 36028797018963936, 72057594037927872, 144115188075855744,  288230376151711488,\newline
       576460752303422976, 1152921504606845952, 2305843009213691904, 4611686018427383808, 9223372036854767616,\newline 18446744073709535232, 36893488147419070464, 73786976294838140928, 147573952589676281856, 295147905179352563712, \newline  590295810358705127424, 1180591620717410254848, 2361183241434820509696, 4722366482869641019392,   9444732965739282038784, \newline
       18889465931478564077568, 37778931862957128155136, 75557863725914256310272, 151115727451828512620544, \newline
       302231454903657025241088, 604462909807314050482176, 1208925819614628100964352, 2417851639229256201928704, \newline
       4835703278458512403857408, 9671406556917024807714816, 19342813113834049615429632, 38685626227668099230859264, \newline
       77371252455336198461718528, 154742504910672396923437056, 309485009821344793846874112, 618970019642689587693748224, \newline
       1237940039285379175387496448, 2475880078570758350774992896, 4951760157141516701549985792, 9903520314283033403099971584, \newline19807040628566066806199943168, 39614081257132133612399886336, 79228162514264267224799772672, \newline 158456325028528534449599545344, 316912650057057068899199090688, 633825300114114137798398181376, \newline1267650600228228275596796362752, 2535301200456456551193592725504, 5070602400912913102387185451008, \newline 5070602400912917605986812821500, 10141204801825835211973625643000, 20282409603651670423947251286000,\newline 40564819207303340847894502572000, 81129638414606681695789005144000, 162259276829213363391578010288000, \newline 324518553658426726783156020576000, 649037107316853453566312041152000, 1298074214633706907132624082304000,\newline 2596148429267413814265248164608000, 5192296858534827628530496329216000, 10384593717069655257060992658432000, \newline 20769187434139310514121985316864000, 41538374868278621028243970633728000, 83076749736557242056487941267456000, \newline166153499473114484112975882534912000, 332306998946228968225951765069824000, 664613997892457936451903530139648000, \newline 1329227995784915872903807060279296000, 2658455991569831745807614120558592000, 5316911983139663491615228241117184000, \newline10633823966279326983230456482234368000, 21267647932558653966460912964468736000, 42535295865117307932921825928937472000, \newline 85070591730234615865843651857874944000, 170141183460469231731687303715749888000, \newline 170141183460469231731687303715884105724, 170141183460469231731687303715884105725.\\
  \hline
\end{tabular}
\end{table}

There also exist several lower bounds which participants referred to.

1) First of all, one could notice that addition chains of length less than 127 cannot produce numbers greater than $2^{126}$. So, we have $\ell(2^{127}-3) > 126$.

2) A more strict bound $\ell(2^{127}-3) > 132$ comes from Sch\"{o}nhage's theorem \cite{chain-3}: $$\ell(n)\geqslant \log_2(n) - \log_2(s(n)) - 2.13,$$ where $s(n)$ denotes the sum of the digits in the binary expansion of $n$.

3) Also, there is the famous Scholz~---~Brauer conjecture \cite{chain-4}: $\ell(2^n-1)\leqslant n-1+\ell(n)$ for any $n\geqslant1$. Moreover, for all $n\leqslant 64$ the inequality becomes the equality as shown in \cite{chain-5}. If we suppose that the conjecture is true and the equality always holds, then it can be assumed that $\ell(2^{127}-1) = 127-1+ 10 = 136$ since $\ell(127) = 10$ \cite{chain-2}. Then, it is easy to see $\ell(2^{127}-1)\leqslant \ell(2^{127}-3) + 1$. Thus, we have that $\ell(2^{127}-3)\geqslant 135$, that is quite close to the shortest found length $136$.

\medskip

At the end, thirteen  teams in the second round were able to find addition chains of length 136 using different approaches, eight team presented chains of length 137 and 138.

\subsection{Problem ``Hash function {\tt FNV2}''}

\subsubsection{Formulation}

The {\tt FNV2} hash function is derived from the function  \href{http://www.isthe.com/chongo/tech/comp/fnv/}{\textcolor{blue}{{\tt FNV-1a}}} \cite{FNV}.
{\tt FNV2} processes a message~$x$ composed of
bytes~$x_1,x_2,\ldots,x_n\in\{0,1,\ldots,255\}$ in the following way:
\begin{enumerate}[noitemsep]
\item[{\bf 1.}]
$h\leftarrow h_0$;
\item[{\bf 2.}]
for $i=1,2,\ldots,n$:
$h\leftarrow (h+x_i)g\bmod 2^{128}$;
\item[{\bf 3.}]
return~$h$.
\end{enumerate}

Here~$h_0=144066263297769815596495629667062367629$ and
$g=2^{88}+315$.

\bigskip

Find a collision, that is, two different messages~$x$ and~$x'$ such
that~$\text{{\tt FNV2}}(x)=\text{{\tt FNV2}}(x')$.
Collisions on short messages and collisions that are obtained without
intensive calculations are welcomed. Supply your answer as a pair of two
hexadecimal strings which encode bytes of colliding messages.

\subsubsection{Solution} We provide a solution based on the Lenstra --- Lenstra --- Lov\'asz (LLL) algorithm. This idea was proposed by several teams.

Firstly, it is clear that $$
\text{{\tt FNV2}}(x_1 x_2\ldots x_n)=
(h_0 g^n + x_1 g^n +x_2 g^{n-1}+\ldots + x_n g) \bmod 2^{128}.
$$

Next, it is sufficient to solve the equation
$$
z_1 g^{n-1} + z_2 g^{n-2} + \ldots + z_n g^0 \equiv 0\!\!\pmod{2^{128}}
$$
in $z_1, z_2,\ldots, z_n \in \{-255,\ldots,255\}$ not equal to zero simultaneously. Indeed, $z_i=x_i - y_i$ for some $x_i, y_i \in \{0, \ldots, 255\}$ and
$$
\text{{\tt FNV2}}(x_1, x_2,\ldots, x_n)- \text{{\tt FNV2}}(y_1, y_2,\ldots, y_n)=
g(z_1 g^{n - 1} + z_2 g^{n-2}+ \ldots + z_n g^0) \equiv 0 \!\!\pmod{2^{128}}.
$$

The purpose is to construct a polynomial such that $g$ is its root. Let us define integer vectors $e^0, \ldots, e^n$ of length $n + 1$ in the following way:
\begin{eqnarray*}
    e^0 &=& (\underbrace{0, \ldots, 0}_n, t \cdot 2^{128}), \text{ where } t \text{ is a small integer},\\
    e^i &=& (\underbrace{0, \ldots, 0}_{i - 1}, 1, \underbrace{0, \ldots, 0}_{n - i}, g^{n - i}\bmod 2^{128}), \text{ where } i \in \{1, \ldots, n\}.
\end{eqnarray*}

Let us add some $z_0$ to $z_1, \ldots, z_n$ and consider the linear combination
$$
    \ell_{z} = z_0 e^0 + \ldots + z_n e^n = (z_1, \ldots, z_n, z_0 t 2^{128} + z_1 g^{n-1} + z_2 g^{n-2} + \ldots + z_n g^0).
$$

To solve the problem it is sufficient to find a linear combination $\ell_{z}$ with $z_1, \ldots, z_n \in \{-255, \ldots, 255\}$ and zero last coordinate.
This can be done using the LLL algorithm. It is a lattice reduction algorithm that can find a short nearly orthogonal basis of $\langle e^0, \ldots, e^n \rangle$. Obtaining such an LLL-reduced basis, we check if it contains a vector $\ell_z$ with desired properties. According to the participants' results, this approach works well starting from $n=17$.

The problem was completely solved by five teams while five more teams provide collisions on long messages. Table \ref{FNV2} contains some collisions proposed by the participants.
\begin{table}[h]
\centering\footnotesize
\caption{Collisions of {\tt FNV2}.}
\smallskip
\label{FNV2}
\begin{tabular}{|c|c|}
\hline
Message 1 & Message 2\\
\hline
\hline
{\tt 808080808080808080808080808080808080} & {\tt a55eca84915f926b4a5f8146c78d8a75d893} \\
\hline
{\tt 8080808080808080808080808080808080} & {\tt c07b375db56d8aceac504381d06696389f} \\
\hline
{\tt 8c2565b0f35411600c3c0e20e21235} & {\tt cb6163c5f3} \\
\hline
``{\tt $\sim\sim\sim\sim$ NSU CRYPTO IS FUN! $\sim\sim\sim\sim$}'' & {\tt 82857b83274c57531e44524e49564b175351273f48572a1c79807c7a} \\
\hline
\end{tabular}
\end{table}

\subsection{Problem ``The image set'' (unsolved)}

\subsubsection{Formulation}

Let $\mathbb{F}_2$ be the finite field with two elements and $n$ be
any positive integer. Let $g(X)$ be an irreducible polynomial of
degree $n$ over $\mathbb{F}_2$. It is widely known that the set of
equivalence classes of polynomials over $\mathbb{F}_2$ modulo $g(X)$
is a finite field of order $2^n$; we denote it by
$\mathbb{F}_{2^n}$.

 Characterize in a non-straightforward way the image set (depending on $n$)
 of the function $F$ over $\mathbb{F}_2^n$ defined as follows:
 $$F(x)=x^3+x.$$ That is, characterize in a way which brings additional information, for instance on its algebraic structure.

\medskip

{\bf An example.} For $n=3$ we can take $g(X)=X^3+X+1$, then each
element of the field $\mathbb{F}_{2^3}$ can be written as a
polynomial of degree at most 2: $a_0+a_1X+a_2X^2$, with
$a_0,a_1,a_2\in \mathbb{F}_2$. We can calculate the table of
multiplication in $\mathbb{F}_{2^3}$ modulo $g(X)$, while the table
of addition just corresponds to adding polynomials over
$\mathbb{F}_{2}$. For example,
$$(1 + X + X^2) + (X + X^2) = 1,$$
$$(X+X^2) (1+X^2) = X+X^2 + X^3+X^4 = 1 + X\!\!\pmod{g(X)}.$$
Now we can calculate all the elements of the image set of $F(x)$.
Indeed, $$\{ F(x)\ |\ x\in \mathbb{F}_{2^3}\} =
\{0,\ 1,\ 1+X,\ 1+X^2,\ 1+X+X^2\}.$$ Then we note that it is the union of
$\{0\}$ and of the affine plane $1+\{0,X,X^2,X+X^2\}$. In our case
it is a desirable algebraic structure of this set.

You need to study this problem for an arbitrary $n$ (or some partial
cases).

\medskip

{\bf Remarks.} Functions over the finite field of order $2^n$ are of
great interest for using in cryptographic applications, for example,
as S-boxes. For instance, AES S-box is based on the inverse function over $\mathbb{F}_{2^8}$. But in fact, there are many open problems in fields of finding new
constructions and descriptions of cryptographically significant
functions!

\subsubsection{Solution}
There were no complete solutions for this problem. Some participants proposed nice ideas. Unfortunately, no one could push them far enough to get significant results. Some did not understand what we were looking for (they focused on the number of solutions, which is known from Mullen et al \cite{MullBook}).

The best solution attempts were proposed by Alexey Udovenko (University of Luxembourg) and by Nikolay Altukhov, Roman Chistiakov, and Evgeniy Manaka team (Bauman Moscow State Technical University). The first solution characterized the case of one pre-image (which is classical), showed a property by the algebraic degree (which gives weak insight on the structure, but it was a nice idea) and finished with observations which are nice but not specific. The second one had an idea of using gcd and tried to calculate it, but did not complete it.

We would also like to recall a known result which may be useful for solving the problem.
\smallskip

\noindent {\bf Theorem.}  \cite{Wil75}
  Let $t_1,t_2$ denote the roots of $t^2+bt+a^3=0$ in $\mathbb{F}_{2^{2n}}$, where $a\in \mathbb{F}_{2^n}, b\in \mathbb{F}_{2^n}^*$.
  Then the factorization of $f(x)=x^3+ax+b$ over $\mathbb{F}_{2^n}$ is characterized as follows:
  \\$\bullet$  $f$ has three zeros in $\mathbb{F}_{2^n}$ if and only if
  $tr_n\left(\frac{a^3}{b^2}+1\right)=0$, where $tr_n$ is the absolute trace function,
  and $t_1,t_2$ are cubes in $\mathbb{F}_{2^n}$ ($n$ even), $\mathbb{F}_{2^{2n}}$ ($n$ odd).
    \\$\bullet$ $f$ has exactly one zero in $\mathbb{F}_{2^n}$ if and only if
  $tr_n\left(\frac{a^3}{b^2}+1\right)=1.$
    \\$\bullet$ $f$ has no zero in $\mathbb{F}_{2^n}$ if and only if
  $tr_n\left(\frac{a^3}{b^2}+1\right)=0$
  and $t_1,t_2$ are not cubes in $\mathbb{F}_{2^n}$ ($n$ even), $\mathbb{F}_{2^{2n}}$ ($n$ odd).

\smallskip

This result depends on $t_1$ and $t_2$ and, when $b\neq 0$, the change of variable $x= bt$ transforms the equation $t^2+bt+a^3=0$ into the equation $x^2+x=\frac{a^3}{b^2}$. So, it may be useful to recall the following fact.

\smallskip

\noindent {\bf Theorem.} \cite{Zin96} Let $n$ be any positive integer and $\beta\in \mathbb{F}_{2^n}$. A necessary and sufficient condition for the existence of solutions in $\mathbb{F}_{2^n}$ of the equation $x^2+x=\beta$ is that $tr_n(\beta)=0$. Assuming that this condition is satisfied, the solutions of the equation are $x=\sum_{j=1}^{n-1}\beta^{2^j}(\sum_{k=0}^{j-1}c^{2^k})$ and $x=1+\sum_{j=1}^{n-1}\beta^{2^j}(\sum_{k=0}^{j-1}c^{2^k})$, where $c$ is any (fixed) element such that $tr_n(c)=1$.

\subsection{Problem ``Boolean hidden shift and quantum computings'' (unsolved)}

\subsubsection{Formulation}

The following long-standing problem is known. Let
$f:\mathbb{F}_2^n\rightarrow\mathbb{F}_2$ be a given Boolean
function. Determine the hidden nonzero shift $a\in\mathbb{F}_2^n$
for the function, i.e. a vector such that $f_a(x)=f(x\oplus a)$ for
all $x\in \mathbb{F}_2^n$. And do it having a limited access to an
oracle for the shifted Boolean function $f_a$ with unknown shift $a$
(i.\,e. a black box, which computes the function $f(a\oplus x)$
for a given vector $x$).
Such a problem is called the {\bf Boolean hidden shift
problem} (BHSP).

In order to solve this problem on a quantum computer, an oracle that
computes the shifted function in the phase is used. This oracle can
be implemented using only one query to an oracle that computes the
function in a register. The phase oracle is a unitary operator defined by its action on the
computational basis:
$O_{f_a}:|x\rangle\mapsto(-1)^{f(x\oplus a)}|x\rangle$, where $|x\rangle$ is the index register.
The {\bf quantum query complexity}  is the minimum number of oracle $O_{f_a}$
accesses needed in the worst case to solve the problem.

There are two classes of Boolean functions for which the quantum
query complexity is minimal and maximal respectively:
\begin{itemize}[noitemsep]
    \item for any bent function, i.\,e. a function in even number of variables that is on the maximal possible Hamming distance from the set of all affine functions, one quantum query suffices to solve the problem exactly~\cite{paper-Quantum-1};
    \item for any delta function, i.\,e. $f(x)=\delta_{x,x_0}$ for some $x_0\in\mathbb{F}_2^n$, the quantum query complexity is $\Theta\left(2^{n/2}\right)$, which is equivalent to Grover's search~\cite{paper-Quantum-2,paper-Quantum-3}.
\end{itemize}

For any Boolean function $f$ in $n$ variables
$
Q(BHSP_f)=O(2^{n/2}),
$
where $Q(BHSP_f)$ is the bounded error quantum query
complexity of the BHSP for~$f$. Moreover, it holds~\cite{paper-Quantum-4}
\begin{equation*}
Q\left(BHSP_f\right)\leqslant\frac{\pi}{4}\frac{2^{n/2}}{\sqrt{{\rm wt}(f)}}+O\left(\sqrt{{\rm wt}(f)}\right),
\end{equation*}
when $1\leqslant {\rm wt}(f)\leqslant 2^{n-1}$, where ${\rm wt}(f)$ is the Hamming weight of $f$.

{\bf The problem to solve is the following:} identify natural
classes of Boolean functions in even number of variables lying
between the two extreme cases of bent and delta functions and
characterize the quantum query complexity of the BHSP for these
functions~\cite{paper-Quantum-4}.

\subsubsection{Solution}
The Boolean hidden shift problem is a particular non-injective case of the well known Hidden Shift problem. There were no complete solutions for this problem. Some attempts to use known results from quantum computation including quantum certificate complexity were made by Andrey Kalachev, Danil Cherepanov, and Alexey Radaev team (Bauman Moscow State Technical University), but no detailed descriptions of classes of Boolean functions with query complexity of the BHSP distinct from two known extremal cases were given.

\subsection{Problem ``Useful Proof-of-work for blockchains'' (unsolved)}

\subsubsection{Formulation}
{\bf Proof-of-work} system is one of the key parts of modern blockchain-based platforms implementations, like cryptocurrency {\bf  Bitcoin} or {\bf Ethereum}.
Proof-of-work means that the user is required to perform some work in order to request some service from the system, e.\,g. to send an e-mail or to create a new block of transactions for the blockchain.

For example, in the Bitcoin system, if some user wants to create a block of transactions and add it to the chain, the hash value of his block must satisfy certain conditions, which can be achieved by iterating special variable $X$ inside the block many times and checking the resulting hash value on every iteration.

What is important about the problem in a proof-of-work system, is that
\begin{itemize}[noitemsep]
    \item{It is known that the solution for the problem exists, and it is also known how many iterations (on average) are required to find it, using best known algorithm $\mathcal{A}$;}
    \item{There are no algorithms for solving the problem, that perform significantly better than $\mathcal{A}$;
    it is believed also that such algorithms will not be found soon;}
    \item{Problem depends on some input data $I$, so you can not find solutions for the problem in advance
    (before input $I$ is known) and then use these solutions without performing any work;}
    \item{Given a problem and a solution to it, it is easy to verify that provided solution is correct.}
\end{itemize}

Unfortunately, solving the problem of finding specific hash values (used in Bitcoin and Ethereum) does not yield any information that is useful outside the system, therefore tremendous amounts of calculations performed to solve the problem are wasted.

Some other implementations of proof-of-work system solve this issue. For example, solutions of proof-of-work problem used in cryptocurrency {\bf Primecoin} give us special chains of prime numbers, useful for scientific research.

{\bf Your task} is to construct a problem $\mathbf{P}$ that can be used in a proof-of-work system, such that information obtained in the proccess of solving it can be useful outside the system.
More formally:
\begin{itemize}[noitemsep]
    \item{$\mathbf{P}$ is, in fact, a family of problems, parametrized by two variables:
    $I$ (input data, you can assume that $I$ is a 256 bit string, or introduce other sensible formats),
    and $C$ (complexity, e.\,g. some positive integer). For fixed input and complexity,
    $\mathbf{P}(I,C)$ is a problem that can be solved by using some algorithm $\mathcal{A}$
    (should be provided in your solution to this task). It should not be possible to find
    a provable solution for the problem $\mathbf{P}(I,C)$ if $I$ is not known;}
    \item{Average time $T$ (amount of computational steps or iterations) required to find a solution of $\mathbf{P}(I,C)$ using algorithm $\mathcal{A}$ is known (assuming input data $I$ is chosen randomly and uniformly), and depends on $C$, so $T=T(C)$, and $T(C)$ can be made very small, infeasibly large, or something in-between by adjusting complexity variable $C$;}
    \item{It should be easy to verify whether any provided solution is correct or not;}
    \item{Any kind of proof that there are likely no significantly better algorithms
    for solving $\mathbf{P}$ than the given algorithm $\mathcal{A}$, is desirable. For example, proof that proposed problem is $\mathcal{NP}$-hard, or any other considerations;}
    \item{You should describe how information obtained in the process of solving $\mathbf{P}$ can be useful outside of the proof-of-work system.}
\end{itemize}

For example, in the Bitcoin system, $\mathbf{P}(I,C)$ is a problem of finding an integer $X$ such that if we apply SHA-256 hash function to the pair $(I,X)$ twice, the resulting hash value, represented as an integer, will not be greater than $C$. Here $C$ is a nonnegative integer, defining complexity of the problem, and $I$ --- a block header, containing information about all transactions included in it, along with some other information --- is an input.

\subsubsection{Solution}
There were no complete solutions for this problem.
Many contestants proposed using NP-hard problems for proof-of-work, but no detailed descriptions how to convert a hard problem into a proof-of-work were provided. In some solutions input data were not linked with the problem. In other solutions the condition of ``easy verifiability'' of the solution to the problem was not satisfied: it is not easy to check an answer ``no'' for an NP-hard decision problem.

An interesting approach for constructing useful proof-of-work was proposed by Carl L{\"o}ndahl (Sweden). Suppose that we ask a user to solve two problems, say $P_1$ and $P_2$. Problem $P_1$ is a regular hash-seeking problem, like the one that is used in Bitcoin. It is easy to link an input with such a problem in order to adjust its difficulty. Problem $P_2$ is some problem based on an NP-hard problem, and $P_2$ input is based on the solution of $P_1$. $P_2$ can have larger variance in time complexity, but we can make average time complexity be dominated by $P_1$, thus keeping overall proof-of-work time consumption more consistent. At the same time, we will be obtaining solutions for some NP-hard problems in the process of obtaining proof-of-work.

\section{Winners of the Olympiad}\label{winners}

\noindent Here we list information about the winners of NSUCRYPTO'2017 (Tables\;\ref{1-sc},\,\ref{1-st},\,\ref{1-pr},\,\ref{2-sc},\,\ref{2-st},\,\ref{2-pr}).

\renewcommand{\topfraction}{0}
\renewcommand{\textfraction}{0}

\begin{figure}[!h]
\centering
\includegraphics[width=0.9\textwidth]{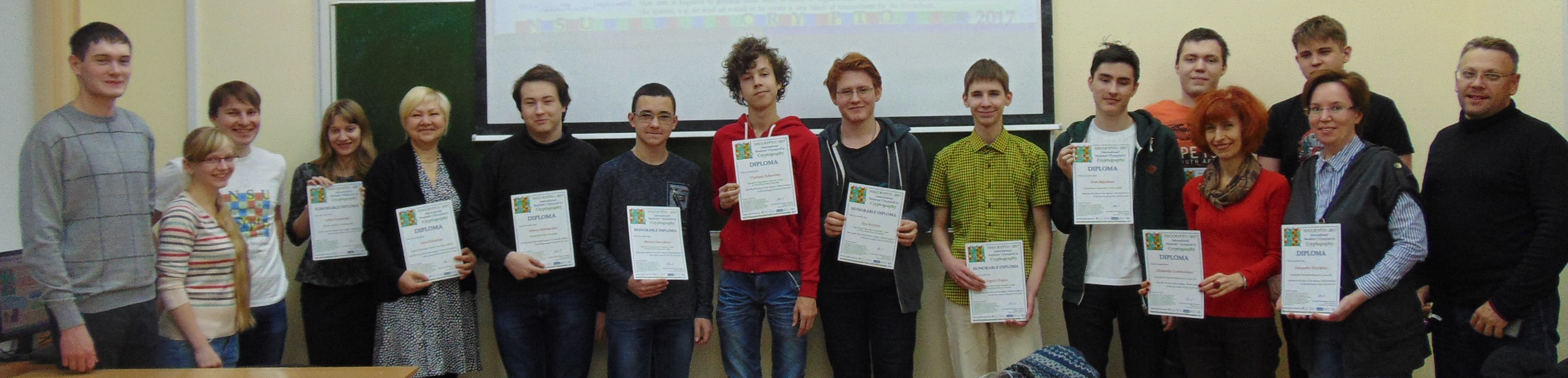}
\vspace{-2mm}
\caption{{\small Awarding ceremony at Novosibirsk State University, December 2017.}}
\label{winners-17}
\end{figure}

\begin{figure}[!h]
\centering
\includegraphics[width=0.9\textwidth]{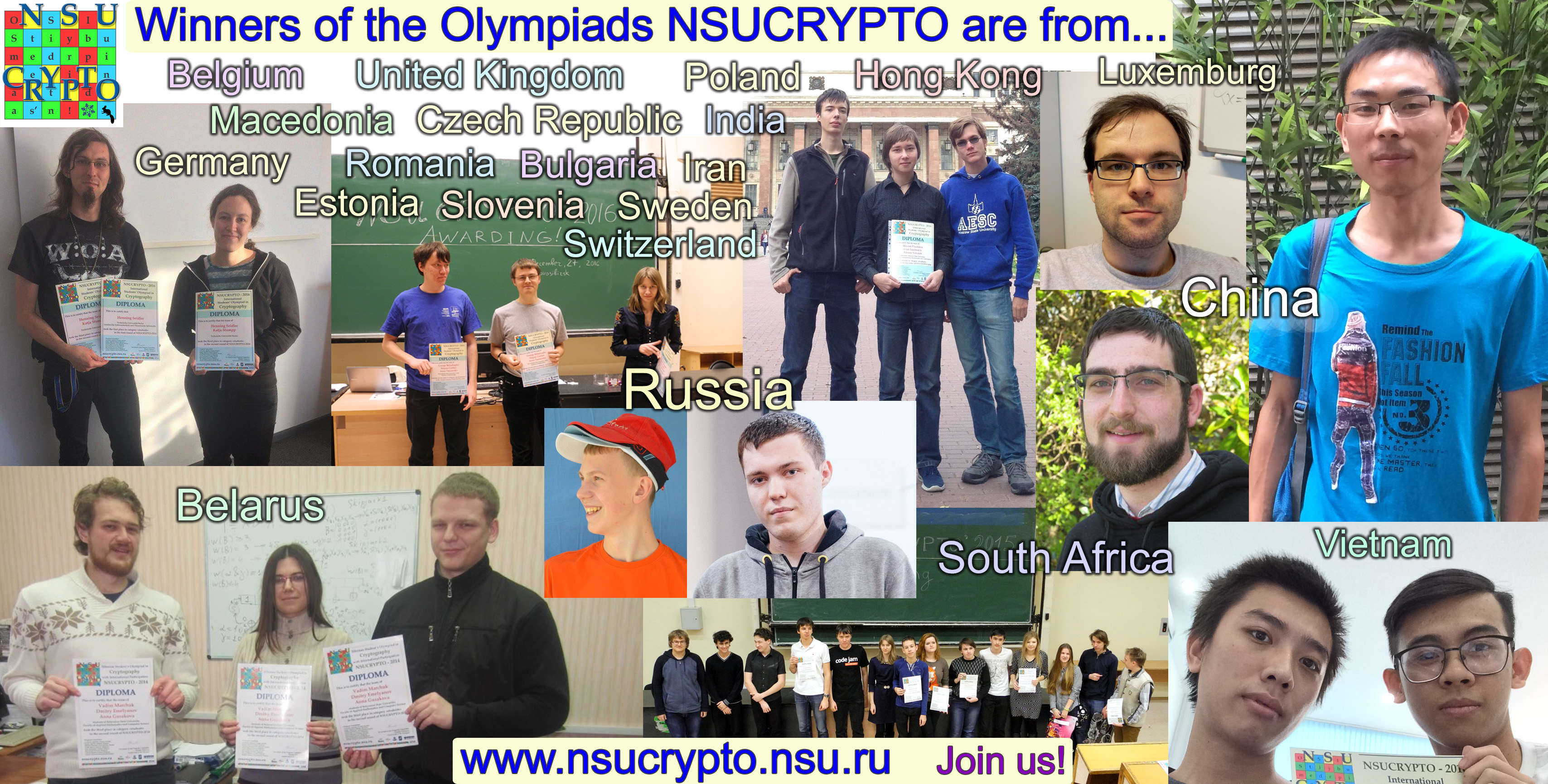}
\vspace{-2mm}
\caption{{\small The NSUCRYPTO winners of different years.}}
\label{winners-all}
\end{figure}

\FloatBarrier

\begin{table}[!h]
\centering\footnotesize
\caption{{\bf Winners of the first round in school section A (``School Student'')}}
\label{1-sc}
\renewcommand{\arraystretch}{1.2}
\renewcommand{\tabcolsep}{1.6mm}
\medskip
\begin{tabular}{|c|l|l|l|c|}
  \hline
  Place & Name & Country, City & School &  Scores \\
  \hline
  \hline
  1 & Alexander Grebennikov & Russia, Saint Petersburg & Presidential PML 239 & 22 \\
   \hline
  1 & Ivan Baksheev & Russia, Novosibirsk   & Gymnasium 6 & 21 \\
  \hline
  2 & Alexander Dorokhin &  Russia, Saint Petersburg    & Presidential PML 239  & 18 \\
  \hline
  3 &   Vladimir Schavelev &    Russia, Novosibirsk & SESC NSU &     17 \\
  \hline
  3 &   Borislav Kirilov &  Bulgaria, Sofia & FPMG &     17 \\
  \hline
  Diploma   &   Ana Kapros  & Romania, Rm Valcea    & National College Mircea cel Batran     &  10 \\
  \hline
  Diploma   &   Filip Dashtevski &  Macedonia, Kumanovo &   Yahya Kemal College & 10 \\
  \hline
  Diploma   &   Andrei Razvan   & Romania, Craiova  &   ``Fratii Buzesti''\newline National College  &   10\\
  \hline
  Diploma   &   Stefan Zaharia &    Romania, Vaslui &       Lyceum Mihail \newline Kogalniceanu     & 9 \\
  \hline
  Diploma   &   Ilia Krytsin &  Russia, Novosibirsk & SESC NSU  & 9 \\
  \hline
  Diploma   &   Grigorii Popov &    Russia, Novosibirsk & SESC NSU  & 8 \\
  \hline
  Diploma   &   Bogdan Circeanu &   Romania, Craiova    & ``Fratii Buzesti''\newline National College    & 8 \\
  \hline
  Diploma   &   Lenart Bucar &  Slovenia, Grosuplje & Gymnasium Bezigrad & 7 \\
  \hline
  Diploma   &   Maxim Desyatkov &   Russia, Kuibyshev   & SESC NSU  & 7 \\
  \hline
\end{tabular}
\end{table}
\begin{table}[!h]
\centering\footnotesize
\caption{{\bf Winners of the first round, section B (in the category ``University Student'')}}
\label{1-st}
\renewcommand{\arraystretch}{1.2}
\renewcommand{\tabcolsep}{0.8mm}
\medskip
\begin{tabular}{|c|l|l|l|c|}
  \hline
  Place & Name  & Country, City & University & Scores \\
  \hline
  \hline
  1 &   Roman Lebedev   & Russia,  Novosibirsk &    Novosibirsk State University    & 20 \\
  \hline
  2 & Robert Spencer & United Kingdom, Cambridge & University of Cambridge  &  15 \\
  \hline
  3 & Nikita Odinokih & Russia,  Novosibirsk    & Novosibirsk State University  &    14\\
  \hline
  3 & Alexey Miloserdov & Russia,  Novosibirsk &    Novosibirsk State University & 14\\
  \hline
  3 & Dheeraj M Pai & India, Chennai    & Indian Institute of Technology, Madras &  13\\
  \hline
  Diploma & Alexey Solovev & Russia, Moscow & Lomonosov Moscow State University & 10\\
  \hline
  Diploma   & Khai Hanh Tang    & Vietnam,  Ho Chi Minh &   University of Science & 9\\
  \hline
  Diploma & Evgeniy Manaka  & Russia, Moscow &  Bauman Moscow State Technical University     &  9\\
  \hline
  Diploma & Andrey Klyuev & Russia, Moscow  & National Research Nuclear University MEPhI  & 8\\
  \hline
  Diploma & Nikolay Altukhov &  Russia, Moscow  & Bauman Moscow State Technical University  & 8\\
  \hline
  Diploma   & Vladimir Bushuev  & Russia, Korolev   & Bauman Moscow State Technical University &    8\\
  \hline
  Diploma & Roman Chistiakov    & Russia, Moscow &  Bauman Moscow State Technical University     & 8\\
  \hline
  Diploma   & Mikhail Sorokin & Russia, Moscow  & National Research Nuclear University MEPhI    & 8\\
  \hline
\end{tabular}
\end{table}
\begin{table}[!h]
\centering\footnotesize
\caption{{\bf Winners of the first round, section B (in the category
``Professional'')}}
\label{1-pr}
\medskip
\renewcommand{\arraystretch}{1.2}
\renewcommand{\tabcolsep}{1.2mm}
\begin{tabular}{|c|l|p{3.8cm}|p{4.0cm}|c|}
\hline
Place & Name & Country, City & Organization & Scores \\
\hline
\hline
1 & Alexey Udovenko &   Luxembourg, Luxembourg  & University of Luxembourg &    28\\
\hline
2   &   Henning Seidler & Germany, Berlin   & TU Berlin &   16\\
\hline
2   & George Beloshapko & Switzerland, Z\"{u}rich   & Google &  15\\
\hline
3   & Daniel Malinowski & Poland, Warsaw &  University of Warsaw    & 12\\
\hline
Diploma &   Evgeniya Ishchukova & Russia, Taganrog  & Southern Federal University   & 8\\
\hline
Diploma &   Egor Kulikov    & Germany, Munich   & dxFeed Solutions GmbH & 8\\
\hline
\end{tabular}
\end{table}


\begin{table}[h]
\centering\footnotesize
\caption{{\bf Winners of the second round (in the category ``School Student'')}}
\label{2-sc}
\renewcommand{\arraystretch}{1.2}
\renewcommand{\tabcolsep}{0.8mm}
\medskip
\begin{tabular}{|c|p{7.5cm}|l|l|c|}
\hline
Place & Names  & Country, City & School & Scores \\
\hline
\hline
Diploma & Filip Dashtevski, Gorazd Dimitrov &   Macedonia, Kumanovo & Yahya Kemal College   & 8\\
\hline
Diploma & Amalia Rebegea, Gabi Tulba-Lecu, Stefan Manolache &   Romania, Bucharest &    CNI ``Tudor Vianu'' & 6\\
\hline
\end{tabular}
\end{table}

\vspace{-0.5cm}

\begin{table}[!h]
\centering\footnotesize
\caption{{\bf Winners of the second round (in the category
``University student'')}}
\label{2-st}
\renewcommand{\arraystretch}{1.2}
\renewcommand{\tabcolsep}{0.8mm}
\medskip
\begin{tabular}{|c|p{5.5cm}|p{2.8cm}|p{4.7cm}|c|}
\hline
Place & Name  & Country, City & University &  Scores \\
\hline
\hline
1 & Roman Lebedev, Vladimir Sitnov,\newline Ilia Koriakin   & Russia, Novosibirsk & Novosibirsk State University &       50 \\
\hline
2   & Alexey Miloserdov, Nikita Odinokih, \newline Saveliy Skresanov & Russia, Novosibirsk  & Novosibirsk State University & 46 \\
\hline
2   & Maxim Plushkin, Ivan Lozinskiy,\newline Azamat Miftakhov & Russia, Moscow & Lomonosov Moscow State\newline  University    &   44\\
\hline
3 & Irina Slonkina & Russia, Moscow &   National Research Nuclear\newline  University MEPhI &  38\\
\hline
3   & Ngoc Ky Nguyen, Thanh Nguyen Van,\newline Phuoc Nguyen Ho Minh    & Vietnam,\newline  Ho Chi Minh City    & Bach Khoa University, Ho Chi\newline Minh University of Technology &      34\\
\hline
3 & Nikolay Altukhov, Roman Chistiakov,\newline Evgeniy Manaka & Russia, Moscow & Bauman Moscow State Technical\newline University  &   32\\
\hline
Diploma & Mikhail Sorokin, Andrey Klyuev,\newline  Anatoli  Makeyev & Russia, Moscow &  National Research Nuclear\newline  University MEPhI &  26\\
\hline
Diploma &   Oskar Soop, Joosep Jääger, Andres Unt   & Estonia, Tartu &  University of  Tartu    &   26\\
\hline
Diploma & Andrey Kalachev, Danil Cherepanov,\newline  Alexey Radaev &   Russia, Moscow  & Bauman Moscow State Technical\newline University  & 24\\
\hline
Diploma & Dianthe Bose &    India, Chennai  & Chennai Mathematical Institute &  23\\
\hline
Diploma & Mikhail Kotov, Oleg Zakharov,\newline  Sergey Batunin & Russia, Tomsk & Tomsk State University &          20\\
\hline
\end{tabular}
\end{table}

\vspace{-0.5cm}

\begin{table}[!h]
\centering\footnotesize
\caption{{\bf Winners of the second round (in the category ``Professional'')}}
\label{2-pr}
\renewcommand{\arraystretch}{1.2}
\renewcommand{\tabcolsep}{1.0mm}
\medskip
\begin{tabular}{|c|p{5.4cm}|p{2.5cm}|p{5.2cm}|c|}
\hline
Place & Names & Country, City & Organization  & Scores \\
\hline
\hline
1   &   Alexey Udovenko &   Luxembourg,\newline Luxembourg  & SnT, University of Luxembourg & 63\\
\hline
2   & Daniel Malinowski, Michal Kowalczyk   & Poland, Warsaw    & University of Warsaw, Dragon Sector   & 49\\
\hline
3   &   Alexey Ripinen, Oleg Smirnov,\newline Peter Razumovsky  & Russia, Saratov    &  Saratov State University &  40 \\
\hline
3   &   Duc Tri Nguyen, Dat Bui Minh Tien,\newline Quan Doan &  Vietnam, \newline Ho Chi Minh city &CERG at George Mason University,\newline Meepwn CTF Team, MeePwn &   37\\
\hline
3   & Carl Londahl &    Sweden, \newline Karlskrona & Independent researcher & 31 \\
\hline
Diploma &   Anna Taranenko  & Russia, \newline Novosibirsk  &   Sobolev Institute of Mathematics    & 29\\
\hline
Diploma &   Kristina Geut, Sergey Titov & Russia, \newline Yekaterinburg &  Ural State University of Railway\newline Transport &    28\\
\hline
Diploma &   Victoria Vlasova, Mikhail Polyakov,\newline Mikhail Tsvetkov    & Russia, Moscow    & Bauman Moscow State Technical\newline University  & 26\\
\hline
Diploma &   Harry Lee, Samuel Tang  & Hong Kong, \newline Hong Kong & Hong Kong University of Science and Technology    & 20\\
\hline
Diploma &Henning Seidler, Katja Stumpp& Germany, Berlin & Berlin Technical University&20\\
\hline
\end{tabular}
\end{table}

\FloatBarrier


\end{document}